\documentclass[10pt]{article} 
\usepackage{fullpage}
\usepackage{graphicx}
\usepackage{upref}
\usepackage{enumerate}
\usepackage{latexsym}
\usepackage{pdfpages}
\usepackage[bookmarks]{hyperref}
\usepackage{color,graphics}
\usepackage{comment} 
\usepackage{caption}
\usepackage{subcaption}
\usepackage{wrapfig}
\usepackage{braket}

\usepackage{amssymb}
\usepackage{amsmath}
\usepackage{amsthm}
\usepackage{mathrsfs}
\usepackage{mathtools}

\usepackage{algorithm,algorithmic,epsfig,graphicx}
\usepackage{amsfonts}
\usepackage{tikz} 
\usepackage{varioref}
\usepackage[ansinew]{inputenc}
\usepackage{qcircuit}
\usepackage{authblk}
\usepackage{multirow}

\theoremstyle{plain}
\newtheorem{theorem}{Theorem}[]
\theoremstyle{plain}
\newtheorem{lemma}{Lemma}[]
\theoremstyle{plain}

\theoremstyle{plain}

\theoremstyle{plain}
\newtheorem{conjecture}{Conjecture}
\theoremstyle{plain}
\newtheorem{claim}{Claim}[]

\theoremstyle{definition}

\newtheorem{definition}{Definition}[]
\theoremstyle{definition}
\newtheorem{fact}{Fact}[]

\theoremstyle{remark}

\theoremstyle{definition}

\AtBeginDocument{}

\DeclareMathOperator{\nat}{\mathbb{N}}

\newcommand{\intg}{\mathbb{Z}}

\newcommand{\poly}{\text{poly}}

\newcommand{\tr}{\text{Tr}}

\newcommand{\id}{\mathbb{I}}

\newcommand{\cliff}{\mathcal{C}}
\newcommand{\pauli}{\mathcal{P}}
\newcommand{\clifft}{\mathcal{J}}

\newcommand{\X}{\text{X}}
\newcommand{\Y}{\text{Y}}
\newcommand{\Z}{\text{Z}}
\newcommand{\had}{\text{H}}
\newcommand{\T}{\text{T}}
\newcommand{\CNOT}{\text{CNOT}}
\newcommand{\phase}{\text{S}}

\newcommand{\chan}[1]{\widehat{#1}}
\newcommand{\tcount}{\mathcal{T}}
\newcommand{\sde}{\text{sde}}
\newcommand{\database}{\mathcal{D}}

\usepackage{multicol}
\usepackage{multirow}

\begin{document}
\title{A polynomial time and space heuristic algorithm for T-count}


\author[1,2,3]{Michele Mosca \thanks{michele.mosca@uwaterloo.ca}}
\author[1,2]{Priyanka Mukhopadhyay \thanks{mukhopadhyay.priyanka@gmail.com, p3mukhop@uwaterloo.ca (corresponding author)}}

\affil[1]{Institute for Quantum Computing, University of Waterloo, Canada}
\affil[2]{Department of Combinatorics and Optimization, University of Waterloo, Canada}
\affil[3]{Perimeter Institute for Theoretical Physics, Waterloo}

\maketitle
\begin{abstract}
An important part of reaping computational advantage from a quantum computer is to reduce the quantum resources needed to implement a desired quantum algorithm. Quantum algorithms that are too large to be practical on noisy intermediate scale quantum (NISQ) devices will require fault-tolerant error correction. This work focuses on reducing the physical cost of implementing quantum algorithms when using the state-of-the-art fault-tolerant quantum error correcting codes, in particular, those for which implementing the $\T$ gate consumes vastly more resources than the other gates in the gate set.

More specifically, in this paper we consider the group of unitaries that can be exactly implemented by a quantum circuit consisting of the Clifford+T gate set. The Clifford+T gate set is a universal gate set and in this group, using state-of-the-art surface codes, the T gate is by far the most expensive component to implement fault-tolerantly. So it is important to minimize the number of T gates necessary for a fault-tolerant implementation. Our primary interest is to compute a circuit for a given $n$-qubit unitary $U$, using the minimum possible number of T gates (called the T-count of unitary $U$). We consider the problem COUNT-T, the optimization version of which aims to find the T-count of $U$. In its decision version the goal is to decide if the T-count is at most some positive integer $m$. Given an oracle for COUNT-T, we can compute a T-count-optimal circuit in time polynomial in the T-count and dimension of $U$. We give a provable classical algorithm that solves COUNT-T (decision) in time $O\left(N^{2(c-1)\lceil\frac{m}{c}\rceil}\poly(m,N)\right)$ and space $O\left(N^{2\lceil\frac{m}{c}\rceil}\poly(m,N)\right)$, where $N=2^n$ and $c\geq 2$. This gives a space-time trade-off for solving this problem with variants of meet-in-the-middle techniques. We also introduce an asymptotically faster multiplication method that shaves a factor of $N^{0.7457}$ off of the overall complexity.

Lastly, beyond our improvements to the rigorous algorithm, we give a heuristic algorithm that outputs a T-count-optimal circuit and has space and time complexity $\poly(m,N)$, under some assumptions. In our heuristic algorithm we developed a novel way of pruning the search space. 
While our heuristic method still scales exponentially with the number of qubits (though with a lower exponent), there is a large improvement by going from exponential to polynomial scaling with $m$.
We implemented our heuristic algorithm with up to 4 qubit unitaries and obtained a significant improvement in time. For all benchmark and random unitaries we studied, the T-count returned by our algorithm is at most the T-count of their circuits shown in previous papers. 
\end{abstract}



\section{Introduction}
\label{sec:intro}

Circuit synthesis and optimization is a significant part of any computer compilation process whose primary goal is to translate from a human readable input (programming language) into instructions that can be executed directly on a hardware. This is true for classical computers and also for quantum computers, which can solve problems believed to be classically intractable, like integer factorization \cite{1994_S,1999_S}. In quantum circuit synthesis the aim is to decompose an arbitrary unitary operation into a sequence of gates from a universal set, which usually consists of Clifford group gates and at least one more non-Clifford gate. Beyond just synthesizing a unitary, we also aim to minimize the overall cost of implementing the corresponding sequence of gates, or circuit, that implements the unitary.

In order to deal with errors due to noise on quantum information, faulty quantum gates, faulty quantum state preparation, faulty measurements, we need a fault-tolerant design, for which a process of quantum error correction is essential.
In particular, for long computations, where the number of operations in the computation vastly exceeds the number of operations one could hope to execute before errors make negligible the likelihood of obtaining a useful answer, fault-tolerant quantum error correction is the only known way to reliably implement the computation. To achieve universality, a non-Clifford gate is required \cite{1998_G, 2004_AG}. As the non-Clifford T gate has known constructions in most of the common error correction schemes, the standard universal fault-tolerant gate set is taken to be ``Clifford+T'' and $\{\had,\T,\CNOT\}$ is a minimal generating set for it.

Most of the popular fault tolerant schemes implement Clifford group gates \emph{transversally}, allowing the logical operations to be performed precisely and with time proportional to the physical gate time. The non-Clifford gates, however, require large ancilla factories and additional operations like gate teleportation and state distillation \cite{2005_BK,1999_GC}. These are less accurate procedures which require both additional time and space compared to a single physical gate. In fact, the cost of fault tolerant implementation of the T gate \cite{2009_FSG,2006_AGP} exceeds the cost of the Clifford group gates by as much as a factor of hundred or more. This is why there are efforts to reduce the T-count for implementations of important unitary operations, such as the work in \cite{2018_G} that manually designed a lower T-count implementation for the specific task of addition. It is also worth noting that the minimum number of T-gates required to implement certain unitaries is a quantifier of difficulty in many algorithms \cite{2016_BG, 2016_BSS} that try to classically simulate quantum computation. Therefore, a very important part of reducing the overall cost of implementing or simulating a unitary is to reduce the number of T gates used.
While alternative fault-tolerance methods such as completely transversal Clifford+T scheme \cite{2013_PR} and anyonic quantum computing \cite{2003_K} are also being explored, minimization of the number of T gates in quantum circuits remain an important and widely studied goal. With recent advances in quantum information processing technologies \cite{2012_BSK_,2011_BWC_,2012_CGC_,2012_RGP_} and fault-tolerant thresholds \cite{2012_BAOKM,2009_FSG,2012_FWH}, as scalable quantum computation is appearing more and more viable \cite{2020_MP} we need efficient automated design tools targeting fault-tolerant quantum computers.

The $n$-qubit unitary implemented by a circuit over the universal Clifford+T gate set is of size $2^n\times 2^n$. There are special classes of circuits like linear reversible circuits which can be described by a matrix of size $n\times n$ \cite{2008_PMH}. But we will not go into their detail, since we are working with general circuits.
A {\bf quantum circuit synthesis algorithm} takes as input an $n$-qubit unitary $U$ of size $2^n\times 2^n$ and outputs a circuit implementing that unitary \cite{2005_DN, 2013_GS, 2013_K}. Often additional constraints are imposed, for example, output a circuit with \emph{minimum} number of certain resources like T gates \cite{2013_GKMR, 2016_dMM}. Let us concentrate on T gates as the resource. These algorithms are often called \textbf{(resource)-optimal circuit synthesis algorithms}. From here on, with a slight abuse of terminology we use the terms \emph{synthesis algorithm} and \emph{T-count-optimal synthesis algorithm} interchangeably. If, at any point they need to be differentiated, it should be clear from the context. We must observe that this puts a tighter constraint on the output of the algorithm and with some good probability may change the complexity of the algorithm by a significant amount. For example, the complexity of synthesizing $n$-qubit quantum circuits is $\poly(N)$, where $N=2^n$ is the input size. But before this work, it was not evident that the complexity of synthesizing T-count-optimal circuits could also be $\poly(N)$. 

In most cases optimal synthesis problems with an input size of $O(2^n)$ becomes too resource-intensive  in practice. So another class of algorithms have been designed, \textbf{quantum re-synthesis algorithms}, where some more information is given about $U$, usually a circuit implementing it \cite{2014_AMMR, 2018_HC}. For re-synthesis algorithms the goal is to \emph{reduce} some resource, like the number of T gates, depending on the input information like the circuit given. 
A single unitary can be implemented by different \emph{equivalent circuits}. For most re-synthesis algorithms the output changes with each different input circuit for the same $U$,  (see for example \cite{2018_AAM}). 

With  unitary of input size $O(2^n)$ we cannot hope to get a resource-optimal synthesis algorithm with complexity less than $O(2^n)$. 
In the literature, nearly every re-synthesis algorithm does not account for the complexity of generating the initial input circuit from $U$. Starting with a general unitary of input size $O(2^n)$ itself has complexity $\poly(2^n)$ \cite{1997_K, 2005_DN}. They report the complexity (in many cases $\poly(n)$) of reducing (not minimizing) a certain resource. A full study comparing the complexity of these two kinds of algorithms and the quality of their results (i.e. how far is the solution of a re-synthesis algorithm from the optimal) is beyond the scope of this paper, and is left for further study.

In spite of their higher complexity compared to re-synthesis algorithms, it is important to study optimal synthesis algorithms. They give optimal solutions and thus can be used for assessing re-synthesis algorithms, for example, how close are their output to an optimal one. From a theoretical point of view, designing synthesis algorithms throws light on the complexity of a problem, which is usually harder than their relaxed re-synthesis counterpart. They can also be useful for re-synthesis algorithms, which usually require a circuit description to begin with. A larger circuit can be fragmented and the unitary of each part can be synthesizd optimally (for example, as done in \cite{2018_MT, 2021_TMK, 2021_GKDetal}). This can yield an entirely different equivalent circuit, which may have fewer T gates (say). In this paper we have designed synthesis algorithms. We will compare the performances of our algorithms with previous synthesis algorithms. 

The Solovay-Kitaev algorithm \cite{1997_K, 2005_DN} guarantees that given a unitary $U$, we can generate a circuit with a universal gate set like Clifford+T, such that the unitary $U'$ implemented by the circuit is at most a certain distance from $U$. In \cite{2013_GS, 2013_KMM} a number of results have been proved by which we can characterize the group of unitaries that can be \emph{exactly} implemented by the Clifford+T set, with or without ancilla. We denote this group by $\clifft_n$. For example, the Toffoli and Fredkin gates belong to $\mathcal{J}_3$. Thus quantum synthesis algorithms can be further sub-divided into two categories : (a) \textbf{exact synthesis algorithms}, that output a circuit implementing $U'=U$ (often up to some global phase) (e.g. \cite{2013_GKMR}) and (b) \textbf{approximate synthesis algorithms}, that output a circuit implementing $U'$ such that $U'$ is \emph{close to} $U$ (e.g. \cite{2016_RS}).

The \textbf{T-count} of unitary $U$ is defined to be the minimum number of $\T$ gates in a Clifford+T circuit that implements it (up to a possible global phase), and is denoted by $\tcount(U)$. In other words, $\tcount(U)$ is the minimum $m$ for which
\begin{eqnarray}
e^{i\phi}U=C_mT_{(q_m)}C_{m-1}T_{(q_{m-1})}\ldots T_{(q_1)}C_0
 \label{eqn:Tcount}
\end{eqnarray}
where $\phi\in[0,2\pi)$, $C_i$ are in the $n$-qubit Clifford group, $q_j\in\{1,2,\ldots,n\}$ and $T_{(r)}$ indicates the T gate acting on the $r^{th}$ qubit. Here we want to emphasize the difference between T-count of a circuit and a unitary. The \textbf{T-count of a circuit} is the number of T-gates in a circuit. The \textbf{T-count of a unitary} is the minimum number of T-gates in any circuit implementing it. Throughout our paper we use the word ``T-count'' to imply T-count of unitary. If we mean T-count of a circuit, it should be clear from the context.
We call a circuit \textbf{T-count-optimal} for $U$ if the number of T gates in the circuit (implementing $U$) is equal to $\tcount(U)$. This implies that among all the circuits that can implement $U$, a T-count-optimal one has the minimum number of T gates.
The T-count of a unitary $U$ may change with the use of additional ancilla qubits and/or measurements with classically controlled operations. For example, Jones \cite{2013_J} has shown how to perform a Toffoli gate using these additional ingredients and only 4 T gates, while the T-count of this unitary without these is $7$ \cite{2013_GKMR}. In our paper ancilla qubits are the only extra resources allowed. 

In this paper we develop \emph{exact synthesis algorithms} for the following problem.

\textbf{MIN-T-SYNTH} : Compute a T-count-optimal quantum circuit for a unitary $U\in\clifft_n$. \\
It has been shown in \cite{2013_GKMR} that an algorithm which computes $\tcount(U)$ can be converted into an algorithm which outputs a T-count-optimal circuit for $U$, with overhead polynomial in $\tcount(U)$ and the dimension of $U$. So we also define the following problem. 

\textbf{COUNT-T} :
Given $U\in\clifft_n$ determine its T-count. \\
The decision version of this problem is obtained by including a positive integer $m$ in the input and an algorithm is required to decide if $\tcount(U)\leq m$.

\subsection{Our results}

We consider the complexity of our exact synthesis algorithms as a function of $m$ and $N=2^n$ and in the RAM model. We treat arithmetic operations on the entries of $U$ at unit cost, and we do not account for the bit-complexity associated with specifying or manipulating them. 

Modifying the algorithm in \cite{2013_GKMR}, we introduce nested meet-in-the-middle technique and obtain a provable algorithm for COUNT-T (decision) having space complexity $O\left(N^{2\lceil\frac{m}{c}\rceil}\poly(m,N)\right)$ (where $c\geq 2$) and time complexity $O\left(N^{2(c-1)\lceil\frac{m}{c}\rceil}\poly(m,N)\right)$. 
The algorithm of \cite{2013_GKMR} has a space and time complexity of $O\left(N^m\poly(m,N)\right)$. Thus our scheme matches the space $\times$ time complexity, and importantly allows for a full space $\times$ time trade-off. Understanding the space-time trade-off of a technique is important to determine the extent to which we can push an algorithm to achieve something in a resource-constrained environment. For example, in \cite{2016_dMM} the authors parallelized the algorithm of \cite{2013_GKMR} with some additional techniques. The unitary with the largest T-count optimal circuit that they could synthesize with 4096 cores was a 4-qubit adder with T-count 7. Reducing the space requirement could have allowed them to synthesize larger circuits. Further details can be found in Section \ref{sec:algo}.

In all our algorithms we work with the channel representation of unitaries (described in Section \ref{prelim:chanRep}) and we are required to perform multiplications of the form $W=\chan{R(P)}V$ a number of times. Here $V,W$ are $N^2\times N^2$ unitary matrices and $\chan{R(P)}$ is the channel representation of $R(P)=\frac{1}{2}\left(1+e^{\frac{i\pi}{4}}\right)\id+\frac{1}{2}\left(1-e^{\frac{i\pi}{4}}\right)P$. Roughly speaking, each $R(P)$ can be implemented with a circuit consisting of only one T gate. More detail has been given in Section \ref{subsec:decompose}. Operator $P$ is an n-qubit non-identity Pauli operator (defined in Section \ref{prelim:cliffPauli}). We developed a fast algorithm in Section \ref{sec:chanRep_n} that computes $W$ in time $O(N^4)$. Currently the fastest algorithm for matrix multiplication has a time complexity $O(N^{4.7457278})$ \cite{2014_lG} for multiplying two $N^2\times N^2$ matrices. This faster matrix multiplication is not only useful for our algorithms, but also may be of independent interest. We have also derived some relations between a certain property of $W$ and $V$ (see Fact \ref{fact:sde} in Section \ref{sec:chanRep_n}). This has played a crucial role in developing our heuristic algorithm in Section \ref{sec:heuristic}. This observation may also be of independent interest and have other applications. 

We give a heuristic algorithm that directly synthesizes a T-count-optimal circuit with both space and time complexity $O\left(\poly(N,\tcount(U))\right)$, thus (partly) answering a question left open in \cite{2013_GKMR}. In our heuristic algorithm we developed a novel way of pruning the search space, depending upon some properties of intermediate unitaries. More details can be found in Section \ref{sec:heuristic}.
To the best of our knowledge, there has been no previous work which hints that complexity of synthesizing T-count-optimal circuits can be polynomial in the input size. This work is focused on methods for optimizing the T-count, while other works have focused on the related task of optimization of T-depth, for example \cite{2013_AMMR, 2013_S}. However, the intuitions developed in this paper have paved the way for further conjectures that have helped in designing a much more efficient T-depth-optimal circuit synthesis algorithm \cite{2021_GMM}, which synthesized previously unknown T-depth-optimal circuits for some standard unitaries like Fredkin, Peres, Negated Toffoli and Quantum XOR. 

In this paper, we have tested our heuristic algorithm on some unitaries whose T-count is known like 3-qubit Toffoli, Fredkin, Peres, Quantum OR, Negated Toffoli gate and 4 qubit 1-bit reversible full adder. In all these cases we obtained a circuit with T-count $7$, thus confirming what was already known \cite{2013_AMMR, 2013_GKMR, 2016_dMM}. But our algorithm is much faster and has a much smaller storage requirement. For example, it gives the T-count-optimal circuit of the 4-qubit adder in about 6 minutes using 1 core. In comparison the current fastest parallel algorithm in \cite{2016_dMM} takes about 12.5 hours with 4096 processors. Here we note that we have a serial algorithm. We also implemented some larger 4-qubit circuits like $U_1=(TOF \otimes \id)(\id \otimes TOF)$ \cite{2014_AMMR} and $U_2=(TOF \otimes \id)(\id \otimes TOF)(TOF \otimes \id)$, where $TOF$ is the 3-qubit Toffoli gate. Had we just plugged in a T-count-optimal circuit for Toffoli we would have obtained a T-count $14$ and 21 respectively. The T-par (re-synthesis) algorithm in \cite{2014_AMMR} obtains a T-count 12 for $U_1$. Instead we obtain a circuit with T-count 11 and 7 respectively in time about 2.2 hours and 6.5 minutes respectively. Of course, the re-synthesis algorithm Tpar is faster (as has been already explained in previous paragraphs), but it is definitely not optimal.
We also tested our algorithms with several random 2, 3 and 4 qubit unitaries formed using 10, 20, 30 and 40 T gates. The T-count of the resulting untiaries may be less than the number of T gates in the circuit used to generate them. So this number serves as an upper bound for the minimum T-count.
Using only the unitaries as input, our algorithm was able to synthesize all of them very rapidly with a T-count either equal to or less than the upper bound  (a non-optimal synthesis algorithm could have produced circuits with T-count exceeding this upper bound). We are not aware of any  rigorous T-count-optimal circuit synthesis algorithms that are able to exactly synthesize unitaries of this size (with a PC). So we cannot confirm the T-count we obtained was optimal, but it is a good indication that our heuristic algorithm might be returning T-count-optimal circuits for most cases.

\subsection{Related work}
\label{subsec:review}

We first review some relevant synthesis algorithms. A nice theoretical understanding about the complexity of synthesizing T-count-optimal circuits has been given in \cite{2019_AM}.
An algorithm for exactly synthesizing unitaries over the Clifford+T gate set was given in \cite{2013_GS} and a superexponentially faster version of this algorithm was given in \cite{2013_K}.
The meet-in-the-middle technique was first used by Amy et al. \cite{2013_AMMR} and Gosset et al. \cite{2013_GKMR} to directly optimize T-depth and T-count respectively, leading to proofs of T-count minimality for various 3 qubit circuits. Building on this technique Di Matteo and Mosca \cite{2016_dMM} constructed a parallel framework using deterministic walks for performing optimal T-count synthesis and confirmed that the 4-qubit 1-bit full adder has optimal T-count 7. The T-count-optimal algorithms in \cite{2013_GKMR,2016_dMM} complexity $O(N^m)$ where $m$ is the T-count. Much work has been done to optimize T-count in exact \cite{2012_KMM,2013_KMM} and approximate \cite{2013_KMM2,2015_S,2016_RS} synthesis of single qubit circuits, as well as repeat-until-success circuits \cite{2014_PS,2015_BRS}. There has been effort to optimize non-Clifford count in two-qubit circuits over other universal fault-tolerant gate set \cite{2020_GRT}.

A lot of work has been done in developing re-synthesis algorithms, for example, \cite{2014_AMMR, 2018_HC, 2019_dBBW, 2019_ZC, 2019_KdW, 2014_ASD}. 

\subsection{Organization}

We give some necessary preliminaries and notations in Section \ref{sec:prelim}. Some results about channel representation of unitaries have been given in Section \ref{sec:chanRep_n}. The provable and heuristic algorithms have been given in Section \ref{sec:algo} and \ref{sec:heuristic} respectively. Finally we conclude in Section \ref{sec:conclude}.

\section{Preliminaries}
\label{sec:prelim}

We write $N=2^n$ and $[K]=\{1,2,\ldots,K\}$. The $(i,j)^{th}$ entry of any matrix $M$ is denoted by $M_{i,j}$ or $M_{ij}$ or $M[i,j]$. We denote the $i^{th}$ row of $M$ by $M[i,.]$ and the $j^{th}$ column by $M[.,j]$. We denote the $n\times n$ identity matrix by $\id_{n\times n}$, $\id_n$ or $\id$ if the dimension is clear from the context.

We call the number of non-zero entries in a matrix as its \textbf{Hamming weight}. We have given some preliminary definitions and facts about the Cliffords, Paulis and the group $\mathcal{J}_n$ generated by Clifford and T gates in Appendix \ref{app:prelim}.

\subsection{Channel representations}
\label{prelim:chanRep}

An $n$-qubit unitary $U$ can be completely determined by considering its action on a Pauli $P_s\in\pauli_n$ : $P_s\rightarrow UP_sU^{\dagger}$. The set of all such operators (with $P_s\in\pauli_n$) completely determines $U$ up to a global phase. Since $\pauli_n$ is a basis for the space of all Hermitian $N\times N$ matrices we can write 
\begin{eqnarray}
 UP_sU^{\dagger} = \sum_{P_r\in\pauli_n} \chan{U}_{rs}P_r,
 \label{eqn:pauliBasis}
\qquad
\text{where}
\qquad
 \chan{U}_{rs} = \frac{1}{2^n}\tr(P_rUP_sU^{\dagger}).
 \label{eqn:chanEntry}
\end{eqnarray}
This defines a $N^2\times N^2$ matrix $\chan{U}$ with rows and columns indexed by Paulis $P_r,P_s\in\pauli_n$. We refer to $\chan{U}$ as the \emph{channel representation} of $U$.

By Hermitian conjugation each entry of the matrix $\chan{U}$ is real. The channel representations respect matrix multiplication i.e. $\chan{UV}=\chan{U}\chan{V}$. Setting $V=U^{\dagger}$ it follows that $\chan{U^{\dagger}}=\big(\chan{U}\big)^{\dagger}$, and we see that the channel representation $\chan{U}$ is unitary.

\begin{lemma}
Let $V$ and $U$ be $N_1\times N_1$ and $N_2\times N_2$ unitaries respectively, where $N_1=2^{m}$ and $N_2=2^{n}$. Then
$
    \chan{(V\otimes U)}=\chan{V}\otimes\chan{U}
$
 \label{lem:chanAncilla2}
\end{lemma}
The proof has been given in Appendix \ref{app:prelim} (Lemma \ref{app:lem:chanAncilla2}).

Let $U'$ is a unitary that acts on the joint state space of ancilla $\ket{\phi}$ and input qubits $\ket{\psi}$. For tensor product inputs, we have
$$
    U'(\ket{\phi}\otimes\ket{\psi})=V\ket{\phi}\otimes U\ket{\psi}=(V\otimes U)(\ket{\phi}\otimes\ket{\psi})
$$
Since the product states span the entire state space, we must have $U'=V\otimes U$. So from Lemma \ref{lem:chanAncilla2} we can calculate $\chan{U'}$ from $\chan{U}$ and $\chan{V}$.
$$
    \chan{U'}=\chan{V}\otimes\chan{U}
$$
Note in many cases the ancilla remains unchanged at the end of operations i.e. $V=\id$.
From here on, with a slight abuse of notation when we write $U\in\clifft_n$ we assume it is the unitary that acts on the joint state space of input and ancilla qubits. 

If $U\in\clifft_n$, implying its entries are in the ring $\intg\left[i,\frac{1}{\sqrt{2}}\right]$, then from Equation \ref{eqn:chanEntry} the entries of $\chan{U}$ are in the same ring. Since $\chan{U}$ is real, its entries are from the subring
$$
    \intg\left[\frac{1}{\sqrt{2}}\right]=\left\{\frac{a+b\sqrt{2}}{\sqrt{2}^k} : a,b\in\intg, \quad k\in\nat \right\}.
$$
The channel representation identifies unitaries which differ by a global phase. We use the following notation for the group of channel representations (where redundant global phases do not appear):
$$
    \chan{\clifft_n}=\left\{\chan{U}:U\in\clifft_n\right\}, \quad \chan{\cliff_n}=\left\{\chan{C}:C\in\cliff_n   \right\}
$$
A matrix $W$ belongs to $\chan{\cliff_n}$ only if it is a unitary matrix with one nonzero entry in each row and each column, equal to $\pm 1$. This is because Cliffords map Paulis to Paulis up to a possible phase factor of $-1$. 

Since the definition of T-count is insensitive to global phase, it is well-defined in the channel representation : for $U\in\chan{\clifft_n}$ we define $\tcount(\chan{U})=\tcount(U)$.
\subsection{Decomposition of unitaries in $\clifft_n$}
\label{subsec:decompose}

In \cite{2013_GKMR} the authors proved a result about the decomposition of unitaries in $\mathcal{J}_n$. 
We can deduce the same conclusion about unitaries which require ancillas (by considering the unitary that acts on the joint state space). 

\begin{theorem}[Proposition 1 in \cite{2013_GKMR} (re-stated)]
 For any $U\in\clifft_n$ there exists a phase $\phi\in[0,2\pi)$, a Clifford $C_0\in\cliff_n$ and Paulis $P_i\in\pauli_n\setminus\{\id\}$ for $i\in[\tcount(U)]$ such that
\begin{eqnarray}
 U=e^{i\phi} \Big(\prod_{i=\tcount(U)}^{1} R(P_i) \Big)C_0
 \qquad
\text{where}\qquad R(P)=\frac{1}{2}(1+e^{\frac{i\pi}{4}})\id+\frac{1}{2}(1-e^{\frac{i\pi}{4}})P.
\end{eqnarray}
\label{thm:decompose}
\end{theorem}

We can also write 
$R(P)=\frac{1}{2}(1+e^{\frac{i\pi}{4}})\id+\frac{1}{2}(1-e^{\frac{i\pi}{4}})CZ_{(q_i)}C^{\dagger}=CT_{(q_i)}C^{\dagger}$, where $C\in\cliff_n$. Using Fact \ref{fact:cliffConj} 
each of these $R(P)$ can be synthesized very efficiently.

\begin{fact}
The channel representation inherits the decomposition from Theorem \ref{thm:decompose} (and in this representation there is no global phase factor.)
\begin{eqnarray}
 \chan{U}=\Big(\prod_{i=\tcount(U)}^{1} \chan{R(P_i)} \Big)\chan{C_0}.
\end{eqnarray}
\label{fact:chan_decompose}
\end{fact}
A unitary $U\in\clifft_n$ can have more than one decomposition as a sequence of $R(P)$s and a Clifford $C_0'\in\cliff_n$, i.e. $U=e^{i\varphi}\left(\prod_{i=t}^1R(P_i)\right)C_0'$. A decomposition in which $t=\tcount(U)$ is called a \textbf{T-count-optimal decomposition}.

\subsection*{Computing T-count-optimal circuits}

A simple application of the decomposition in Theorem \ref{thm:decompose} is the following result shown in \cite{2013_GKMR}: 

\emph{Suppose $\mathcal{A}$ is an algorithm which solves the decision problem COUNT-T. For any $U\in\clifft_n$, with overhead polynomial in $N$ and $\tcount(U)$, such an algorithm can also be used to generate a T-count-optimal circuit for $U$ over the gate set $\{\had,\T,\CNOT\}$. 
}

\subsection*{T-count for single qubit unitaries}

For single qubit unitaries $U$ it has been shown in \cite{2013_GKMR} that the T-count can be directly computed from its channel representation $\chan{U}$. For this the authors introduced the following quantity and proved the following results.
\begin{definition}
 For any non-zero $v\in\intg\left[\frac{1}{\sqrt{2}}\right]$ the \textbf{smallest denominator exponent}, denoted by $\sde(v)$, is the smallest $k\in\nat$ for which
 $$
    v=\frac{a+b\sqrt{2}}{\sqrt{2}^k} \qquad \text{ with } a,b\in\intg.
 $$
 We define $\sde(0)=0$. By the above definition $a$ is odd when $k>0$.
 For a $d_1\times d_2$ matrix $M$ with entries over this ring we define 
 $$
    \sde(M)=\max_{a\in[d_1],b\in[d_2]} \sde(M_{ab}).
 $$
\end{definition}

\begin{fact}[Fact 2 in \cite{2013_GKMR}]
 Let $q,r\in\intg\left[\frac{1}{\sqrt{2}}\right]$ with $\sde(q)>\sde(r)$. Then 
 $$
    \sde\left(\frac{1}{\sqrt{2}}(q\pm r)\right)=\sde(q)+1.
 $$
 \label{fact:sde_uneq}
\end{fact}

\begin{fact}
 Let $q,r\in\intg\left[\frac{1}{\sqrt{2}}\right]$ with 
 $\sde(q)=\sde(r)>0$. Then 
 $$
    \sde\left(\frac{1}{\sqrt{2}}(q\pm r)\right)\leq\sde(q).
 $$
 \label{fact:sde_eq}
\end{fact}

\begin{proof}
Let $q=\frac{a+b\sqrt{2}}{\sqrt{2}^k}$ and $r=\frac{c+d\sqrt{2}}{\sqrt{2}^k}$ such that $k>0$ and $a,c\in 2\intg+1$. Since $(a\pm c)\in 2\intg$ we have
\begin{eqnarray}
    \frac{1}{\sqrt{2}}(q\pm r)&=&\frac{1}{\sqrt{2}^{k+1}}\left((a\pm c)+(b\pm d)\sqrt{2} \right)
        = \frac{1}{\sqrt{2}^k}\left((b\pm d)+\left(\frac{a\pm c}{2}\right)\sqrt{2} \right). \nonumber
\end{eqnarray}
If $(b\pm d)\in 2\intg$ we can reduce the denominator exponent further. The fact follows.
\end{proof}

\begin{theorem}[Theorem 1 in \cite{2013_GKMR}]
 The T-count of a single-qubit unitary $U$ is :
 $
    \tcount(U)=\sde(\chan{U}).
 $
\end{theorem}
But the above characterization does not generalize to $n>1$ qubits. As a counter-example  the sde of the channel representation of the Toffoli gate is 2 but its T-count is 7.

\section{Properties of $\chan{R(P)}$}
\label{sec:chanRep_n}

In this section, we make some observations about the structure of the channel representation, $\chan{R(P)}$ where $P\in\pauli_n$. Since $\chan{\id_{n\times n}}=\id_{n^2\times n^2}$, so we focus on the non-identity Paulis. This leads to more compact representation (and hence less storage space) and efficient computation of these unitaries. It also gives much faster algorithms for certain operations like multiplication by these unitaries (or their inverses). This reduces the overhead (space and time complexity) in algorithms involving these operations.  

Let $U=R(P)=\frac{1}{2}(1+e^{\frac{i\pi}{4}})\id+\frac{1}{2}(1-e^{\frac{i\pi}{4}})P$ where $P\in\pauli_n\setminus\{\id\}$. 
So $U^{\dagger}=\frac{1}{2}(1+e^{\frac{-i\pi}{4}})\id+\frac{1}{2}(1-e^{\frac{-i\pi}{4}})P$, since $P^{\dagger}=P$. Thus
\begin{eqnarray}
 P_rUP_sU^{\dagger}
 &=&\frac{1}{4}(2+\sqrt{2})P_rP_s+\frac{i\sqrt{2}}{4}P_rP_sP 
 -\frac{i\sqrt{2}}{4}P_rPP_s+\frac{1}{4}(2-\sqrt{2})P_rPP_sP.
 \label{eqn:P_rUP_sU+}  \nonumber
\end{eqnarray}

Let $P=\bigotimes_{i=1}^nA_i$, $P_r=\bigotimes_{i=1}^nB_i$ and $P_s=\bigotimes_{i=1}^nC_i$ where $A_i,B_i,C_i\in\{\id,\X,\Y,\Z\}$. Hence
\begin{eqnarray}
 K_{rs}'&=&\tr(P_rUP_sU^{\dagger})=\frac{1}{4}(2+\sqrt{2})\prod_{j=1}^n\tr(B_jC_j) 
 +\frac{i\sqrt{2}}{4}\prod_{j=1}^n\tr(B_jC_jA_j) \nonumber\\
&&-\frac{i\sqrt{2}}{4}\prod_{j=1}^n\tr(B_jA_jC_j) 
 +\frac{1}{4}(2-\sqrt{2})\prod_{j=1}^n\tr(B_jA_jC_jA_j)   \nonumber
 \label{eqn:K'}
\end{eqnarray}
and $\chan{U}_{rs}=\frac{1}{2^n}K_{rs}'$. 

We note the following fact.
\begin{fact}
For any $P\in\pauli_n$ if $P=\bigotimes_{j=1}^nA_j$ where $A_j\in\{\id,\X,\Y,\Z\}$, then $\tr(P)=\prod_{j=1}^n \tr(A_j)=0$ if there exists any $A_j\neq\id$.
 \label{fact:tr0}
\end{fact}

Let $\chan{U}$ is indexed by Paulis $P_r,P_s\in\pauli_n$ in its rows and columns and both the first row and column has index $\id^{\otimes n}$. 

\begin{claim}
The first row and column has all $0$, except the first entry at $\chan{U}_{\id^{\otimes n},\id^{\otimes n}}$, which is $1$.
 \label{claim:row1}
\end{claim}

\begin{proof}
For the first entry we have $P_r=P_s=\id^{\otimes n}$. 
 So $K_{rs}=\chan{U}_{rs}=\frac{1}{2^n}\tr(UU^{\dagger})=\frac{1}{2^n}\cdot 2^n=1$.
 
 Now take the first row $P_r=\id^{\otimes n}$ i.e. $B_j=\id$ $\forall j=1,2,\ldots,n$. Consider any other entry except the first one, so $C_j\neq\id$ for some $j$. Then
 \begin{eqnarray}
  K_{rs}'&=&\frac{1}{4}(2+\sqrt{2})\prod_{j=1}^n\tr(C_j)+\frac{i\sqrt{2}}{4}\prod_{j=1}^n\tr(C_jA_j) 
 -\frac{i\sqrt{2}}{4}\prod_{j=1}^n\tr(A_jC_j)+\frac{1}{4}(2-\sqrt{2})\prod_{j=1}^n\tr(A_jC_jA_j) \nonumber \\
 &=& \frac{1}{4}(2+\sqrt{2})\prod_{j=1}^n\tr(C_j)   
 + \left( \frac{i\sqrt{2}}{4}\prod_{j=1}^n\tr(C_jA_j) 
 -\frac{i\sqrt{2}}{4}\prod_{j=1}^n\tr(C_jA_j) \right) \nonumber \\
 &&+\frac{1}{4}(2-\sqrt{2})\prod_{j=1}^n\tr(A_jA_jC_j) \nonumber \\
 &=& 0 \qquad [\text{Fact } \ref{fact:tr0} \text{ and since } A_j^2=\id ]. \nonumber
 \end{eqnarray}
 The second equality follows by using the cyclic property of trace.
 
 A similar argument follows for first column where $P_s=\id^{\otimes n}$ i.e. $C_j=\id\quad\forall j=1,2,\ldots,n$.

\end{proof}

\subsection*{Diagonal elements of $\chan{U}$}

Let us look at the diagonal elements when $r=s$, so $P_r=P_s$ (or equivalently $B_j=C_j\quad\forall j$). So
\begin{eqnarray}
 \chan{U}_{rr}&=&\frac{1}{4}(2+\sqrt{2})\prod_{j=1}^n\tr(\id)   
 +\frac{1}{4}(2-\sqrt{2})\prod_{j=1}^n\tr\left((B_jA_j)^2\right).
 \label{eqn:K'diag}
\end{eqnarray}

\begin{claim}
The diagonal entries of $\chan{U}$ are $1$ or $\frac{1}{\sqrt{2}}$. 

 \label{claim:diagonal}
\end{claim}

\begin{proof}
From Equation (\ref{eqn:K'diag}) we have:
\begin{eqnarray}
 K_{rr}'&=&\frac{1}{4}(2+\sqrt{2})\prod_{j=1}^n\tr(\id)+\frac{1}{4}(2-\sqrt{2})\prod_{j=1}^n\tr(\pm\id) \nonumber \\
 &=&2^n \text{ or } \frac{2^n}{\sqrt{2}} \nonumber\\
 &&\text{(depending on whether there are an even or } 
 \text{odd number of $\tr(-\id)$ terms in the product).} \nonumber
\end{eqnarray}

Since $\chan{U}_{rr}=\frac{1}{2^n}K_{rr}'$, the claim follows.
\end{proof}

\paragraph{\emph{Further observation :}}
From Claim \ref{claim:diagonal} (and its proof) we can say that a diagonal entry is $\frac{1}{\sqrt{2}}$ if the second summand has odd number of multiplicands with $-1$. This happens if for an odd number of $j$ we have $A_j\neq B_j$ and $A_j,B_j\in\pauli_1\setminus\{\id\}$ (and $C_j=B_j$).  

\subsection*{Off-diagonal elements of $\chan{U}$ }

In an off-diagonal entry we have $r\neq s$, so $\exists k\in[n]$ such that $B_k\neq C_k$. So $(B_kC_k), (B_kA_kC_kA_k) \in\pauli_1\setminus\{\id\}$ and $\tr(B_kC_k)=\tr(B_kA_kC_kA_k)=0$. Thus
\begin{eqnarray}
 K_{rs}'&=&\frac{i\sqrt{2}}{4}\prod_{j=1}^n\tr(B_jC_jA_j)   
 -\frac{i\sqrt{2}}{4}\prod_{j=1}^n\tr(C_jB_jA_j) \quad [r\neq s]
 \label{eqn:K'offDiag}
\end{eqnarray}

\begin{claim}
If $r\neq s$ (equivalently if $\exists k\in[n]$ such that $B_k\neq C_k$) then $\chan{U}_{rs}=\pm\frac{1}{\sqrt{2}}$ if and only if $B_jC_j=\pm i A_j$ for an odd number of $j\in [n]$ and for the rest of the co-ordinates we have $B_jC_jA_j=C_jB_jA_j=\id$ (i.e. $B_j=C_j$ and $A_j=\id$).
 \label{claim:offDiagonal}
\end{claim}

\begin{proof}
If the given condition is satisfied then from Equation \ref{eqn:K'offDiag} we have:
\begin{eqnarray}
 K_{rs}'&=&(\pm i)^m\frac{i\sqrt{2}}{4}\prod_{j=1}^n\tr(\id)-(\mp i)^m\frac{i\sqrt{2}}{4}\prod_{j=1}^n\tr(\id) 
 = \pm \frac{2^n}{\sqrt{2}} \qquad [m \text{ is odd}]    \nonumber
\end{eqnarray}

In the other direction suppose $K_{rs}'=\pm\frac{2^n}{\sqrt{2}}$. In this case  we must have $(B_jC_jA_j),(C_jB_jA_j)=\id$ or $\pm i\id$ for all $j=1,2,\ldots,n$. From Equation \ref{eqn:K'offDiag} we can conclude that there should exist odd number of $j$ such that $\tr(B_jC_jA_j)=-\tr(C_jB_jA_j)$. This is possible if for an odd number of $j$ we have\\ $B_jC_j=\pm iA_j$. 

Thus the claim follows.
\end{proof}

\paragraph{Correlation between the row and column entries}
We note the following correlation between row and column entries.

\begin{claim}
\begin{enumerate}
    \item If a diagonal entry is $1$ then all other entries in the corresponding row and column is $0$.
    \item If a diagonal entry is $\frac{1}{\sqrt{2}}$ then one other entry in the corresponding row is $\pm \frac{1}{\sqrt{2}}$ and one other entry in the corresponding column is $\mp\frac{1}{\sqrt{2}}$.
 \end{enumerate}
  \label{claim:diagOffdiag}
\end{claim}
The proof has been given in Appendix \ref{app:RP} (Claim \ref{app:claim:diagOffdiag})
\begin{claim}
 Exactly $2^{2n-1}$ diagonal elements can be $\frac{1}{\sqrt{2}}$.
 \label{claim:diagSign}
 \end{claim}
 The proof has been given in Appendix \ref{app:RP} (Claim \ref{app:claim:diagSign}).

\subsection{A compact representation of $\chan{R(P)}$}
\label{subsec:compactChanRep}

From the above discussions we can deduce that it is sufficient to represent the $N^2\times N^2$ matrix $\chan{R(P)}$ as an array of length $N^2/4$ i.e. $2^{2n-2}$. 

Let $R(P)=U$ for some $P\in\pauli_n\setminus\{\id\}$. Assume in row $i$ we have $\chan{U}_{ii}=\frac{1}{\sqrt{2}}$ and $\chan{U}_{i\ell}=\pm\frac{1}{\sqrt{2}}$. By Claim \ref{claim:diagOffdiag} we can say that $\chan{U}_{\ell\ell}=\frac{1}{\sqrt{2}}$ and $\chan{U}_{\ell i}=\mp\frac{1}{\sqrt{2}}$. All other entries in these two rows are $0$.

So we can store all the necessary information in an array $A_{\chan{U}}$ where each entry of this array is a pair of the form $(i,\pm \ell)$ ($i<\ell$) which signifies that $\chan{U}_{ii}=\frac{1}{\sqrt{2}}$ and $\chan{U}_{i\ell}=\pm\frac{1}{\sqrt{2}}$. This implies $\chan{U}_{\ell\ell}=\frac{1}{\sqrt{2}}$ and $\chan{U}_{\ell i}=\mp\frac{1}{\sqrt{2}}$. We need not store $(\ell,\mp i)$. And for values $k$ where $U_{kk}=1$, none of the stored pairs of integers will contain $\pm k$. This array $A_{\chan{U}}$ is of size $N^2/4$.

\subsection{Multiplication of $\chan{R(P)}$ with any matrix}
\label{subsec:multChanRep}

Let $U=R(P)$ where $P\in\pauli_n\setminus\{\id\}$. We want to analyze the multiplication of $\chan{U}$ with some matrix $V$ of dimension $N^2\times N^2$. Let $W=\chan{U}V$.
\begin{eqnarray}
 W_{ij} = \sum_{k=1}^{N^2} \chan{U}_{ik}V_{kj}
 \label{eqn:mult}
\end{eqnarray}

Let us consider the following cases.

 \textbf{Case 1 :} \textbf{Diagonal element of $\chan{R(P)}$ is $1$ }
 
 Let $\chan{U}_{ii}=1$. Then we know from Claim \ref{claim:diagOffdiag} that $\chan{U}_{ik}=0$ for all $k\neq i$. So
 \begin{eqnarray}
  W_{ij} = V_{ij} \qquad \forall j\in[N^2]    \nonumber
 \end{eqnarray}
Thus $W[i,.]=V[i,.]$, in other words, the $i^{th}$ row of $V$ gets copied into the $i^{th}$ row of the product matrix $W$.

\textbf{Case 2 :} \textbf{Diagonal element of $\chan{R(P)}$ is $\frac{1}{\sqrt{2}}$ }

Let $\chan{U}_{ii}=\frac{1}{\sqrt{2}}$. Then we know from Claim \ref{claim:diagOffdiag} that $\exists \ell\in[N^2]\setminus\{i\}$ such that $\chan{U}_{i\ell}=\pm\frac{1}{\sqrt{2}}$ and $\chan{U}_{\ell i}=\mp\frac{1}{\sqrt{2}}$. Also $\chan{U}_{\ell \ell}=\frac{1}{\sqrt{2}}$. Thus
$$
    W_{ij} = \chan{U}_{ii}V_{ij}+\chan{U}_{i\ell}V_{\ell j}=\frac{1}{\sqrt{2}}(V_{ij}\pm V_{\ell j}) \quad
    \forall j\in[N^2].
$$
So $W[i,.]=\frac{1}{\sqrt{2}}[V[i,.]\pm V[\ell,.]]$. Similarly
$$
    W_{\ell j} = \chan{U}_{\ell\ell}V_{\ell j}+\chan{U}_{\ell i}V_{ij}=\frac{1}{\sqrt{2}}(V_{\ell j}\mp V_{ij}) \quad\forall j\in[N^2]
$$
which gives $W[\ell,.]=\frac{1}{\sqrt{2}}[V[\ell,.]\mp V[i,.]]$. Thus the $i^{th}$ and $\ell^{th}$ row of $W$ becomes a linear combination of the $i^{th}$ and $\ell^{th}$ rows of $V$.

To summarize, using the compact representation ($A_{\chan{U}}$) of $\chan{U}=\chan{R(P)}$ we can construct the product $W$ using the following algorithm (MULT).
\begin{enumerate}
 \item For every $j=1,\ldots,N^2/4$ do the following:
        \begin{enumerate}
         \item Let $(i,\pm\ell)\leftarrow A_{\chan{U}}[j]$.
         \item $W[i,.]\leftarrow\frac{1}{\sqrt{2}}\Big[V[i,.]\pm V[\ell,.] \Big]$.
         \item $W[\ell,.]\leftarrow\frac{1}{\sqrt{2}}\Big[V[\ell,.]\mp V[i,.] \Big]$.
         \item $S\leftarrow S\bigcup \{i,\ell\}$. ($S$ is a set to store the indices of non-unity diagonal elements.)
        \end{enumerate}
 \item For every $k\in[N^2]\setminus S$
    \begin{enumerate}
     \item $W[k,.]\leftarrow V[k,.]$.
    \end{enumerate}
\end{enumerate}

So this provides reduced space and time complexity for the computation of channel representation of any unitary in $\clifft_n$, provided we know its decomposition into product of $R(P)$.

\subsection*{Inverse of $\chan{R(P)}$}

The inverse of $\chan{R(P)}$ can be calculated by the following algorithm (INV).

Let $A_{\chan{U}}$ be the compact array representation of $\chan{U}=\chan{R(P)}$. Then the compact representation of its inverse $\chan{U^{-1}}$ is given by the array $A_{\chan{U^{-1}}}$ and can be calculated in the following way.

\begin{enumerate}
 \item For every $j=1,\ldots,N^2/4$ do the following:
    \begin{enumerate}
     \item Let $(i,\pm \ell)\leftarrow A_{\chan{U}}[j]$.
     \item $A_{\chan{U^{-1}}}\leftarrow (i,\mp \ell)$.
    \end{enumerate}
\end{enumerate}

\textbf{\emph{Correctness}}

Let $V$ is the matrix whose compact representation is given by $A_{\chan{U^{-1}}}$. It is sufficient if we prove $\chan{U}V=W=\id$.

Note the rows which have a single one in the diagonal is same for both $\chan{U}$ and $V$ and hence in the product matrix $W$ these rows remain the same.

Now consider the rows which have $\frac{1}{\sqrt{2}}$ in the diagonal of $\chan{U}$. Let $A_{\chan{U}}[j]=(i,\ell)$. By the above algorithm INV $A_{\chan{U^{-1}}}[j]=(i,-\ell)$. This implies $V_{ii}=V_{\ell\ell}=\frac{1}{\sqrt{2}}$, $V_{i\ell}=-\frac{1}{\sqrt{2}}$ and $V_{\ell i}=\frac{1}{\sqrt{2}}$. Also $\chan{U}_{ii}=\chan{U}_{\ell\ell}=\frac{1}{\sqrt{2}}$, $\chan{U}_{i\ell}=\frac{1}{\sqrt{2}}$ and $\chan{U}_{\ell i}=-\frac{1}{\sqrt{2}}$.

Using the algorithm MULT we have $W[i,.]=\frac{1}{\sqrt{2}}\left[V[i,.]+V[\ell,.]\right]$ and $W[\ell,.]=\frac{1}{\sqrt{2}}\left[V[\ell,.]-V[i,.]\right]$. Thus $W_{ii}=W_{\ell\ell}=1$ and $W_{i\ell}=W_{\ell i}=0$. The remaining entries in these two rows are all $0$.

Hence we can conclude that $W=\id$.

\subsection*{Complexity of the multiplication algorithm MULT}

Let $W=\chan{U}V$ and $U=R(P)$, where $P\in\pauli_n$. Note the unitaries are of dimension $N^2\times N^2$. From the above algorithm we see that half of the rows of $V$ get copied to $W$. For the remaining $N^2/2$ rows we perform component-wise addition or subtraction 
and multiplication among pairs of rows of $V$. Thus we have the following result.
\begin{theorem}
 $W=\chan{R(P)}V$ (where $P\in\pauli_n$) can be computed in time $O\left(\frac{N^4}{2}\right)$. 
 \label{thm:multRP}
\end{theorem}

Currently the fastest algorithm for matrix multiplication has a time complexity $O(N^{4.7457278})$ \cite{2014_lG} for multiplying two $N^2\times N^2$ matrices. So the algorithm MULT can be very useful as the dimension $N=2^n$ increases or in situations where we have to do many such multiplications.

\subsection{A data structure for the channel representation}

We give a data structure that will avoid working with floating point arithmetic. This became specially useful while implementing the heuristic algorithm given in Section \ref{sec:heuristic}, where it is crucial we keep track of the sde of a unitary matrix.

The entries of the channel representation $\chan{U}$ are in the ring $\intg\left[\frac{1}{\sqrt{2}}\right]$. Thus if $v=\chan{U}[i,j]$ is an entry then it can be written as $v=\frac{a+b\sqrt{2}}{\sqrt{2}^k}$, where $a,b\in\intg$ and $k\in\nat$. We can store $v$ as a tuplet $[a,b,k]$. We make sure that $k$ is the sde of $v$. Thus every time $v$ gets modified we make necessary reductions. If $a$ is odd or $k=0$ it cannot be reduced any further. If $a$ is even then we perform the following procedure: \\
While $a$ is even and $k>0$:
\begin{eqnarray}
 (1) b'\leftarrow b,\quad (2) a\leftarrow a/2,\quad (3) b\leftarrow a,\quad (4)a\leftarrow b'. \nonumber
\end{eqnarray}

\subsection*{sde of product matrix}

Here we make some further observations about the sde of the product matrix while multiplying by $\chan{R(P)}$ ($P\in\pauli_n$). This again becomes useful for the heuristic algorithm in Section \ref{sec:heuristic}. 

Let $W=\chan{R(P)}V$, where $V$ is some unitary. We look at one particular entry $W_{ij}$ and following algorithm MULT assume $W_{ij}=\frac{1}{\sqrt{2}}(q+r)$. From Fact \ref{fact:sde_uneq} we can deduce that $\sde(W_{ij})=\sde(V_{ij})+1$ if $\sde(q)\neq\sde(r)$ and hence $\sde(W)$ maybe $\sde(V)+1$. But if $\sde(q)=\sde(r)$ then from Fact \ref{fact:sde_eq}\\
$\sde(W_{ij})\leq\sde(V_{ij})$. So $\sde(W)$ maybe less than or equal to $\sde(V)$. 

Note that if we apply $\chan{R(P)^{-1}}$ to $W$ then the resultant should have $\sde(V)$. We know sde cannot increase by more than $1$. Thus we can conclude that:
\begin{fact}
$\sde(W)=\sde(V)\pm 1$ or $\sde(W)=\sde(V)$.
\label{fact:sde}
\end{fact}
 In the special case of single-qubit unitaries \\
 $\sde(W)=\sde(V)+1$ ($P\neq\id$). This fact results in efficient algorithms for exact synthesis of T-count-optimal single-qubit unitaries.

\section{A provable algorithm for COUNT-T }
\label{sec:algo}

In this section we modify and extend the meet-in-the-middle (MITM) algorithm in \cite{2013_GKMR}, and achieve a provable algorithm for COUNT-T (decision) that can be implemented with much less space while retaining the overall space $\times$ time complexity. The space complexity of our algorithm is $O\left(N^{\frac{2m}{c}}\poly(m,N)\right)$ while the time complexity is $O\left(N^{\frac{2(c-1)m}{c}} \poly(m,N) \right)$ ($c\geq 2$). This kind of algorithm becomes especially useful if we want to store the database and run the algorithm a number of times to test the T-count of a number of unitaries. Reducing the space requirement may also allow one to synthesize larger circuits. For example, in \cite{2016_dMM} the authors parallelized the framework in \cite{2013_GKMR} with some additional techniques that reduced the search time, though asymptotically the complexities remained same. But they did not reduce the space requirement. So the largest circuit that they could synthesize with 4096 processors was the 4-qubit adder with T-count 7.  

Suppose we wish to decide if $\tcount(U)\leq m$ where $U\in\clifft_n$. From Theorem \ref{thm:decompose} we know that any such unitary can be expressed as $e^{i\phi}\left(\prod_{i=m}^1 R(P_i) \right)C_0$, for some phase $\phi$. Utilising this result, we can search over expressions of the form $U^{\dag}\prod_{i=m}^1 R(P_i)$ until we find one that is a global phase times an element of the Clifford group $\cliff_n$.

\begin{figure}[h]
\centering
\includegraphics[width=7.5cm, height=3cm]{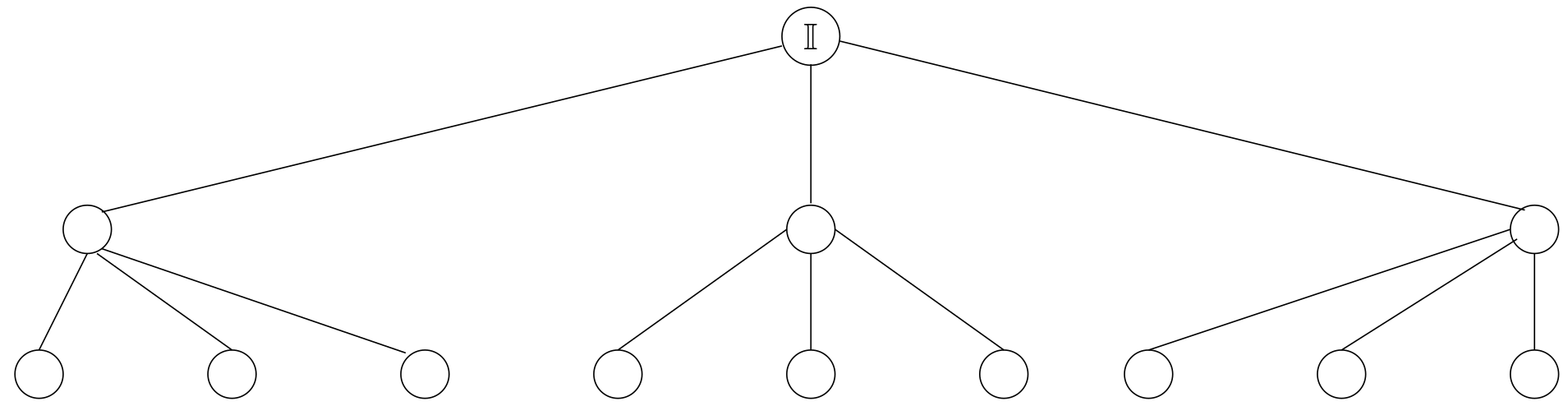}
\caption{Part of the procedure of the provable algorithm can be compared to building a tree up to a certain depth. The root of the tree stores $\id$.}
 \label{fig:provable}
\end{figure}

It might be useful to compare this procedure with building a tree, whose nodes store some unitary and the edges represent some $R(P)$. Thus the unitary in a child node is obtained by multiplying the unitary of the parent node with the $R(P)$ represented by the edge. Thus there can be at most $N^2-1$ children nodes of any parent node, since $|\pauli_n\setminus\{\id\}|=N^2-1$. The root in this case stores $\id$ and we assume it is at depth $0$ (Figure \ref{fig:provable}).

In an exhaustive search we might have to build this tree till depth $m$ and then search for an unitary $V$ in this tree such that $U^{\dag}V$ is a global phase times an element of $\cliff_n$. This gives a search space of size at most $N^{2m}$. To reduce the search space and hence time, we use a slightly more complicated procedure. Very concisely, we build this tree till depth $\lceil\frac{m}{c}\rceil$, where $c \geq 2$. If we do not find some appropriate unitary within this depth then with the help of the unitaries in the leaves we perform a recursive search in the remaining tree. In the meet-in-the-middle search algorithm of \cite{2013_GKMR} $c=2$. We modify this and take $c>2$. We call it the ``nested meet-in-the-middle algorithm'' because of the recursive application of meet-in-the-middle method.

We provide some necessary definitions and results from \cite{2013_GKMR} that will be useful in describing our algorithm. We work with the channel representation (the group $\chan{\clifft_n}$) and consider the left cosets of $\chan{\cliff_n}$ in $\chan{\clifft_n}$. The following notion determines whether two unitaries are from the same coset. 

\begin{definition}[\textbf{Coset label}]
Let $W\in\chan{\clifft_n}$. Its coset label $W^{(co)}$ is the matrix obtained by the following procedure.
(1) Rewrite $W$ so that each nonzero entry has a common denominator, equal to $\sqrt{2}^{\sde(W)}$. (2) For each column of $W$, look at the first non-zero entry (from top to bottom) which we write as $v=\frac{a+b\sqrt{2}}{\sqrt{2}^{\sde(W)}}$. If $a<0$, or if $a=0$ and $b<0$, multiply every element of the column by $-1$. Otherwise, if $a>0$, or $a=0$ and $b>0$, do nothing and move on to the next column. (3) After performing this step on all columns, permute the columns so they are ordered lexicographically from left to right.

 \label{defn:cosetLabel}
\end{definition}

Then the following can be shown.
\begin{theorem}[\textbf{Proposition 2 in \cite{2013_GKMR}}] 
Let
$W,V\in\chan{\clifft_n}$. Then $W^{(co)}=V^{(co)}$ if and only if 
$W=VC$ for some $C\in\chan{\cliff_n}$.
 \label{thm:cosetLabel}
\end{theorem}

Using the coset labels we can construct a \emph{sorted coset database} as follows.
\begin{definition}[\textbf{Sorted coset database $\database_k^n$}]
For any $k\in\nat$, a sorted coset database $\database_k^n$ is a list of unitaries $W\in\chan{\clifft_n}$ with the following three properties:
\begin{enumerate}
 \item Every unitary in the database has T-count $k$. In other words, every $W\in\database_k^n$ satisfies
 $\tcount(W)=k$.
 \item For any unitary with T-count $k$, there is a unique unitary in the database with the same coset label. For any $V\in\chan{\clifft_n}$ with $\tcount(V)=k$, there exists a unique $W\in\database_k^n$ such that $W^{(co)}=V^{(co)}$.
 \item The database is sorted according to the coset labels. If $W,V\in\database_k^n$ and $W^{(co)}<V^{(co)}$ (using lexicographic ordering on the matrices) then $W$ appears before $V$.
\end{enumerate}

 \label{defn:sortCoset}
\end{definition}

\subsection{Nested MITM algorithm}
\label{subsec:nestMeet}

We build a sorted coset database of unitaries with T-count at most $\lceil\frac{m}{c}\rceil$, where $c>2$ i.e. $\database_{\lceil\frac{m}{c}\rceil}^n$. If we cannot find the coset label of the given unitary $U$ within this database, it implies its T-count is greater than $\lceil\frac{m}{c}\rceil$. Then we check whether the T-count is at most $2\lceil\frac{m}{c}\rceil$ by a meet-in-the-middle search, using $\database_{\lceil\frac{m}{c}\rceil}^n$. 
If we do not find a match at this step then using $\database_{\lceil\frac{m}{c}\rceil}^n$ we perform a ``sort of'' meet-in-the-``middle'' search and check if the T-count is at most $3\lceil\frac{m}{c}\rceil$. We carry on these iterations till we have tested for T-count at most $m$. If our search has been unsuccessful in all the previous steps then we conclude that T-count of the given unitary $U$ is greater than $m$. 

A more detailed description of the algorithm is given below.

\textbf{Input : } (i) A unitary $U\in\clifft_n$ and (ii) a non-negative integer $m$.

\textbf{Output : } $\tcount(U)$ if it is less than $m$ or return NO.

\begin{enumerate}
 \item[0.] We calculate the channel representation $\chan{U}$ and its coset label $\chan{U}^{(co)}$.

 \item \textbf{Precompute sorted coset databases.}   \\
 $\database_0^n, \database_1^n,\ldots,\database_{\lceil\frac{m}{c}\rceil}^n$, for positive integer $c>2$.
 
 We start with $\database_0^n$ which contains only the 
 $N^2\times N^2$ identity matrix. Then $\database_1^n,\database_2^n,\ldots,\database_{\lceil\frac{m}{c}\rceil}^n$ are constructed recursively as follows. To construct $\database_k^n$ we consider all unitaries of the form
 \begin{eqnarray}
  W=\chan{R(P)}M 
  \label{eqn:database}
 \end{eqnarray}
where $M\in\database_{k-1}^n$ and $P\in\pauli_n\setminus\{\id\}$ sequentially. We insert $W$ into $\database_k^n$ (maintaining the ordering according to the coset labels) if and only if its coset label is new.
 
 \item \textbf{Check if } $\tcount(U)\leq\lceil\frac{m}{c}\rceil$. 
 
 Use binary search to check if there exists $W\in\database_j^n$ for some $j\in\{0,1,\ldots,\lceil\frac{m}{c}\rceil\}$, such that $\chan{U}^{(co)}=W^{(co)}$. If so, OUTPUT $\tcount(U)=j$ and stop. Else, CONTINUE to step 3.
 
 \item \textbf{Nested meet-in-the-middle}.
 
 Set $j=1$. \\
 While $j\lceil\frac{m}{c}\rceil < m$ do the following:
    \begin{enumerate}
        \item Let $B=\min\{(j+1)\lceil\frac{m}{c}\rceil,m\}$.
        
        \item \textbf{Check if } $j\lceil\frac{m}{c}\rceil <\tcount(U)\leq B$.
        
        Let $r=B-j\lceil\frac{m}{c}\rceil$. For each $W\in\database_r^n$ we calculate $V=W^{\dag}\chan{U}$. We recurse through the steps done before to check if $\tcount(V) \leq j\lceil\frac{m}{c}\rceil$. If it returns $i$ then we OUTPUT $\tcount(\chan{U})=\tcount(V)+r$ and STOP.
        Else, we set $j\leftarrow j+1$ and REPEAT the while loop.
     
    \end{enumerate}

 \item OUTPUT NO implying $\tcount(U)>m$.
 
\end{enumerate}

\subsection*{Complexity} 
\label{dummy}

We calculate the time and space complexity of the above algorithm assuming the RAM model. To compute the sorted coset database we loop over all unitaries in Equation \ref{eqn:database}, with $k\in\{0,1,\ldots,\lceil\frac{m}{c}\rceil\}$. There are $O\left(N^{2\lceil\frac{m}{c}\rceil}\right)$ such unitaries. For each unitary we need to compute the coset label and search to find it in the database generated so far. This takes time $O\left(\log\left(N^{2\lceil\frac{m}{c}\rceil}\right)\right)$, since the database is sorted. The unitaries and their coset labels are of size $N^2\times N^2$. So step 1 takes time $O\left(N^{2\lceil\frac{m}{c}\rceil}\poly(m,N) \right)$. To store the database similarly requires space 
$O\left(N^{2\lceil\frac{m}{c}\rceil}\poly(m,N) \right)$. Step 2 takes time $\poly(m,N)$ for binary search through the sorted database. 

Now let us consider the complexity of the while loop in step 3. Consider the first iteration. Let $t=\lceil\frac{m}{c}\rceil$. There are $O\left(N^{2\lceil\frac{m}{c}\rceil}\right)$ unitaries in $\database_{\lceil\frac{m}{c}\rceil}^n$ and so $O\left(N^{2\lceil\frac{m}{c}\rceil}\right)$ number of $V$ are calculated serially in step 3b. For each such $V$, to check if $\tcount(V)$ is at most $\lceil\frac{m}{c}\rceil$ requires time $\poly(m,N)$. Thus the first iteration of while loop takes time $O\left(N^{2\lceil\frac{m}{c}\rceil}\poly(m,N)\right)$. We assume time taken to complete $(j-1)^{th}$ iteration is in $O\left(N^{2(j-1)\lceil\frac{m}{c}\rceil}\poly(m,N)\right)$. Now let us consider the $j^{th}$ iteration. Again we compute at most $O\left(N^{2\lceil\frac{m}{c}\rceil}\right)$ unitaries $V$ at step 3b. For each $V$, to test if their T-count is at most $j\lceil\frac{m}{c}\rceil$ takes time $O\left(N^{2(j-1)\lceil\frac{m}{c}\rceil}\poly(m,N)\right)$. So total time taken to complete the $j^{th}$ iteration is $O\left(N^{2j\lceil\frac{m}{c}\rceil}\poly(m,N)\right)$.

Thus by induction we can conclude that the total time taken at step 3 is at most 
$O\left(N^{2(c-1)\lceil\frac{m}{c}\rceil}\poly(m,N)\right)$. Hence the time complexity of the given algorithm is 
$O\left(N^{2(c-1)\lceil\frac{m}{c}\rceil}\poly(m,N)\right)$ and space complexity is $O\left(N^{2\lceil\frac{m}{c}\rceil}\poly(m,N)\right)$.

\subsection*{Correctness}

The algorithm correctly generates the sorted coset databases $\database_0^n,\database_1^n,\ldots,\database_{\lceil\frac{m}{c}\rceil}^n$. The proof is similar to the one given in \cite{2013_GKMR} (Section 5.2) and so we skip it here. Given this fact it follows that if $0\leq\tcount(U)\leq\lceil\frac{m}{c}\rceil$ then step 2 of our algorithm will correctly compute $\tcount(U)$.

We now prove that if $\lceil\frac{m}{c}\rceil<\tcount(U)\leq m$ then our algorithm correctly computes it at step 3. It is sufficient to show that this property holds for any particular iteration of the while loop. That is, if $j\lceil\frac{m}{c}\rceil<\tcount(U)\leq\min\{(j+1)\lceil\frac{m}{c}\rceil,m\}$ then the algorithm correctly returns it in the $j^{th}$ iteration. Our proof is similar to Theorem 2 in \cite{2013_GKMR}. To prove the property holds in the first iteration we can clearly re-state the result of \cite{2013_GKMR} as follows.
\begin{theorem}[\textbf{Theorem 2 in \cite{2013_GKMR} (re-stated) }]
Let $U\in\clifft_n$ and $m\in\nat$ with $\lceil\frac{m}{c}\rceil<\tcount(U)\leq B$, where $B=\min\{2\lceil\frac{m}{c}\rceil,m\}$, and let $\database_0^n,\ldots,\database_{\lceil\frac{m}{c}\rceil}^n$ be sorted coset databases. Then $t=\tcount(U)$ is the smallest integer in $\left\{\lceil\frac{m}{c}\rceil+1,\lceil\frac{m}{c}\rceil+2,\ldots,B\right\}$ for which
$
    \left(W^{\dag}\chan{U}\right)^{(co)}=V^{(co)}
$
for some $W\in\database_{B-\lceil\frac{m}{c}\rceil}^n$ and $V\in\database_{t-B+\lceil\frac{m}{c}\rceil}^n$.
 \label{thm:GKMRthm2}
\end{theorem}
Now we assume that this property holds till the $(j-1)^{th}$ iteration. In the next theorem we prove that it remains true in the next iteration.
\begin{theorem}
Let $m\in\nat$, $c\geq 2$ and $j\geq 1$ are integers. $U\in\clifft_n$ is a unitary with $j\lceil\frac{m}{c}\rceil<\tcount(U)\leq B$, where $B=\min\{(j+1)\lceil\frac{m}{c}\rceil,m\}$. Let $\database_0^n,\database_1^n,\ldots,\database_{\lceil\frac{m}{c}\rceil}^n$ are sorted coset databases.

Then $t=\tcount(U)$ is the smallest integer in $\{j\lceil\frac{m}{c}\rceil+1,j\lceil\frac{m}{c}\rceil+2,\ldots,B\}$ for which
\begin{eqnarray}
 \left(W^{\dag}\chan{U} \right)=V
 \label{eqn:thmNestMeet}
\end{eqnarray}
for some $W$ with $W^{(co)}\in\database_{r}^n$, where $r=B-j\lceil\frac{m}{c}\rceil$, and $V$ is a unitary such that $\tcount(V)\leq j\lceil\frac{m}{c}\rceil$, and $t=r+\tcount(V)$.
 \label{thm:nestMeet}
\end{theorem}
The proof of this theorem has been given in Appendix \ref{app:provable} (Theorem \ref{app:thm:nestMeet})

If none of the iterations of the while loop in step 3 return $\tcount(U)$ then it implies $\tcount(U)>m$ and the algorithm returns NO.

This completes the proof of correctness of our algorithm.

\section{Polynomial complexity heuristic algorithm}
\label{sec:heuristic}

In this section we describe a heuristic algorithm that returns a T-count-optimal circuit in polynomial time and space. We developed a novel way of pruning the search space. We group the intermediate unitaries according to some properties and then select one such group according to some criteria. None of the previous algorithms use pruning, nor was there any evidence that an efficient pruning of the search space is possible for this particular problem. More detail explanation has been given in the following paragraphs.

\begin{figure}[h]
\centering
\includegraphics[width=7.5cm, height=3cm]{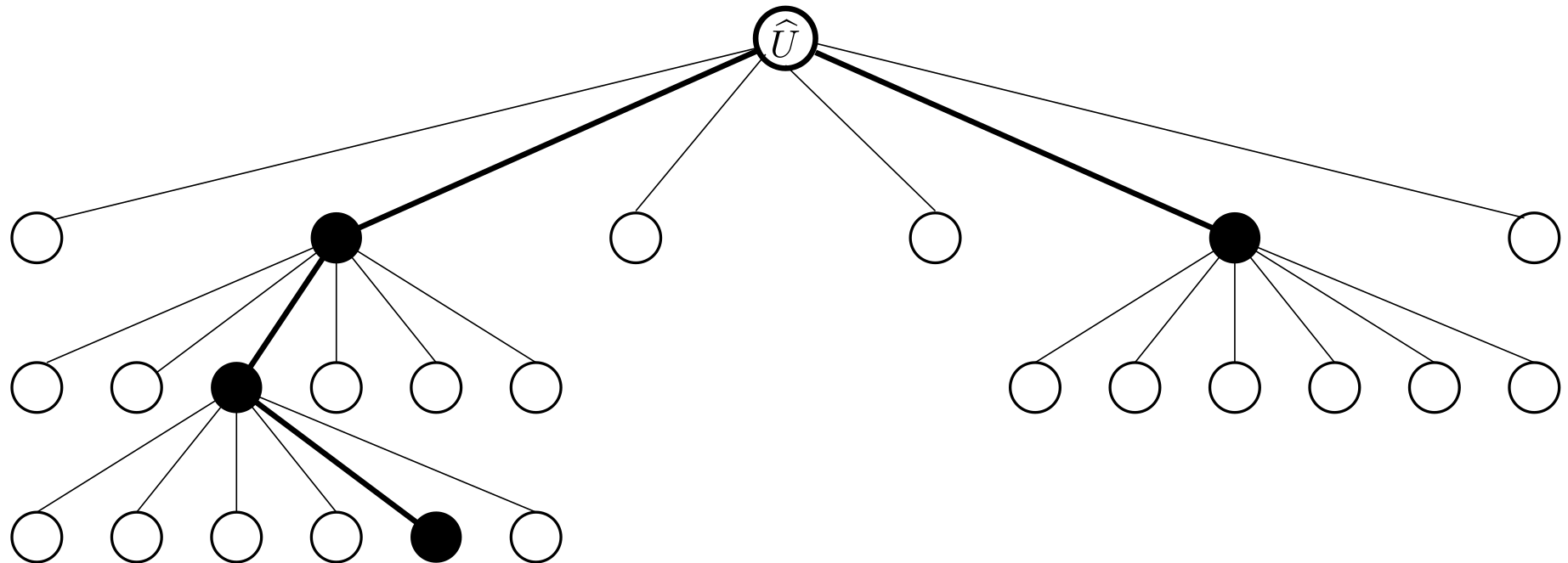}
\caption{The tree built in the heuristic procedure. At each level we select a set of nodes according to some changes in the properties of the child unitaries with respect to their parents, for example change in sde and Hamming weight. Unitaries in the next level are generated from the selected set (black nodes). We stop building the tree as soon as we reach a unitary with sde 0. The path length to this node (in this case 3) is the T-count of the unitary $U$. It also gives us the optimal decomposition of $U$.
}
 \label{fig:heuristic}
\end{figure}

In the provable algorithm described in Section \ref{sec:algo} we used a nested meet-in-the-middle technique to search for a set of $\chan{R(P_i)}$ such that 
$\chan{U}^{\dag}\prod_{i=\tcount(U)}^1\chan{R(P_i)}$ is $\chan{C_0}$ for some $C_0\in\cliff_n$.
Alternatively, we can also search for a set of $\chan{R(P_i)}^{-1}$ such that $\left(\prod_{i=1}^m\chan{R(P_i)}^{-1}\right)\chan{U}$ is $\chan{C_0}$ for some $C_0\in\cliff_n$, which is the approach taken by our heuristic algorithm. As in Section \ref{sec:algo} we build a tree but instead we store $\chan{U}$ in the root. This time the edges represent $\chan{R(P)}^{-1}$ for some $P\in\pauli_n\setminus\{\id\}$. We stop building this tree the moment we reach a node which stores $\chan{C_0}$ for some $C_0\in\cliff_n$. This implies that the path from the root to this leaf gives a decomposition of $\chan{U}$. Unlike Section \ref{sec:algo}, we build a number of trees with increasing depth. We stop at the first tree which gives us a decomposition. In other words, in our synthesis algorithm MIN-T-SYNTH, we perform a number of iterations of a procedure $\mathcal{A}$, in which we build one such tree with target depth increasing by 1 in each iteration.
In these kinds of searches the size of the tree is one of the important factors that determine the complexity of the algorithm. To reduce the complexity we try to prune this tree (Figure \ref{fig:heuristic}).

At each level we try to group the nodes according to some ``properties'' or ``parameters'' of the unitaries stored in them. We hope that these parameters will ``distinguish'' the ``correct'' nodes at each level or depth of the tree and thus we would get a decomposition. Note there can be more than one decomposition of $\chan{U}$ with the same or different T-count. By ``correct'' nodes we mean those that occur in a T-count-optimal decomposition  of $\chan{U}$. If the parameters always select only the correct nodes then we expect to get much fewer nodes at each level of the tree and the number of levels we have to build is $\tcount(U)$. But the parameters we selected did not always distinguish the correct nodes and there were some false positives. In order for the algorithm to succeed we have to be careful so that we do not lose all correct nodes at any level and to make it efficient we have to ensure that the number of false positives are not too large and they eventually get eliminated.

We selected two parameters - sde and Hamming weight of the unitaries. We know from Fact \ref{fact:sde} that sde of a child node unitary can differ by at most 1 from its parent node unitary. While building a unitary we start with $\chan{\id_n}$ and multiply by subsequent $\chan{R(P_i)}$ till we reach $\chan{U}$. We have observed that in most of these multiplications the sde increases by 1 and the Hamming weight also gradually increases until it (Hamming weight) reaches the maximum. So while doing the inverse operations i.e. decomposing $\chan{U}$ we expect that in most of the steps sde will decrease by 1 and as we get close to the identity, the Hamming weight will also likely decrease. If we multiply by a ``wrong'' $\chan{R(P)}^{-1}$ we expect to see the same changes with much less probability, which is the probability of the false positives. This helps us distinguishe the ``correct'' and ``wrong'' nodes.

Specifically, at each level we divide the set $S$ of nodes into some subsets and select one of them. Below are three possible ways to divide the nodes that we have found effective. Suppose in one instance of $\mathcal{A}$, $m$ is the target depth (maximum depth of the tree to be built) and we have built the tree till depth $i$.
\begin{enumerate}
 \item[A.] We divide into two sets - $S_0$ (sde increase) and $S_1$ (sde decrease). We select the set with the minimum cardinality such that the sde of the unitaries in this set is at most $m-i$. We include the nodes with unchanged sde in it. 
 
 \item[B.] We divide into 4 sets - $S_{00}$ (both sde and Hamming weight increase), $S_{01}$ (sde increase, Hamming weight decrease), $S_{10}$ (sde decrease, Hamming weight increase) and $S_{11}$ (both sde and Hamming weight decrease). Nodes with unchanged Hamming weight but sde increase are put in both $S_{00}$ and $S_{01}$, while nodes with unchanged Hamming weight but sde decrease are put in both $S_{10}$ and $S_{11}$. We select the set with the minimum cardinality such that the sde of the unitaries in this set is at most $m-i$. We include in it the nodes with unchanged sde (irrespective of the change in Hamming weight).
 
 \item[C.] We divide into 9 sets - $S_{00}$ (both sde and Hamming weight increase), $S_{01}$ (sde increase but Hamming weight decrease), $S_{02}$ (sde increase but Hamming weight same), $S_{10}$ (sde decrease, Hamming weight increase) and $S_{11}$ (both sde and Hamming weight decrease), $S_{12}$ (sde decrease but Hamming weight same), $S_{20}$ (sde same but Hamming weight increase), $S_{21}$ (sde same but Hamming weight decrease), $S_{22}$ (both sde and Hamming weight same). We select the set with the minimum cardinality such that the sde of the unitaries in this set is at most $m-i$.
\end{enumerate}
We follow any one of the above methods of divide-and-select throughout the algorithm. Note in each of the above methods, if the cardinality of the unitaries in the selected set is more than $m-i$ then it implies we cannot get sde $0$ nodes within the next few levels.

The algorithm MIN-T-SYNTH is as follows:

\textbf{Input : } $\chan{U}$ such that $U\in\clifft_n$.

\textbf{Output : } $[P_{\tau(U)}, P_{\tau(U)-1},\ldots, P_1, \chan{C_0}]$ such that $\chan{U}=\left(\prod_{i=\tcount(U)}^1 \chan{R(P_i)}\right)\chan{C_0}$ where $C_0\in\cliff_n$.

\begin{enumerate}
 \item $m=\sde(\chan{U})$.

 \item While(1)
    \begin{enumerate}
     \item Call $\mathcal{A}(\chan{U},m)$ (described below).
     
     \item If returns a decomposition then
                return the decomposition and
                break.
      \item Else $m\leftarrow m+1$.      
    \end{enumerate}

\end{enumerate}

A description of sub-routine $\mathcal{A}$ is as follows.

\textbf{Input : } (i) $\chan{U}$ such that $U\in\clifft_n$,
(ii) a non-negative integer $m$.

\textbf{Output : } If a decomposition exists then 
$[P_{m'}, P_{m'-1},\ldots,1, \chan{C_0} ]$ such that $m'\leq m$ and 
$\chan{U}=\left(\prod_{i=m'}^1 \chan{R(P_i)}\right)\chan{C_0}$ where $C_0\in\cliff_n$, else NO.

\begin{enumerate}
 \item  $Path\leftarrow []$; $\widetilde{U}\leftarrow \{(\chan{U},Path)\}$. \\
 /* $Path$ is an array that stores the path (or sequence of $\chan{R(P)}^{-1}$) to the current node. */
 
 \item For $i=1,2,\ldots,m$ do the following:
    \begin{enumerate}
     \item Multiply each unitary in $\widetilde{U}$ by $\chan{R(P_i)}^{-1}$, for $P_i\in\pauli_n\setminus\{\id\}$.
     
     \item Note the changes in $\sde$ and Hamming weight. If sde of any product unitary is $0$ then STOP and return the $Path$ till this node and the unitary itself ($\chan{C_0}$).
     
     \item Else, apply a divide-and-select method (A, B or C, which should be followed throughout the algorithm) described before. Note (by the above discussion) the selected set $S_{min}$ (say) has unitaries with sde at most $m-i$. If no such set is found then STOP and return NO. \\
     /* $\sde > m-i$ implies we cannot decrease it to $0$ within the next $m-i$ iterations.*/
     
     \item Empty $\widetilde{U}$ and include the product unitaries in $S_{min}$ along with their $Path$.
    \end{enumerate}
\end{enumerate}

\paragraph*{Space and time complexity} The analysis of space and time complexity of the algorithm MIN-T-SYNTH is based on the following assumptions. 
\begin{conjecture}
\begin{enumerate}
    \item[A.] The cardinality of the set $\widetilde{U}$ in each iteration of $\widetilde{\mathcal{A}}$ is at most $\poly(N)$, when method A of divide-and-select is applied. 
    
    \item[B.] The cardinality of the set $\widetilde{U}$ in each iteration of $\widetilde{\mathcal{A}}$ is at most $\poly(N)$, when method B of divide-and-select is applied.
    
    \item[C.] The cardinality of the set $\widetilde{U}$ in each iteration of $\widetilde{\mathcal{A}}$ is at most $\poly(N)$, when method C of divide-and-select is applied.
    
    \item[D.] We get at least one T-count-optimal decomposition.
\end{enumerate}
\label{conj:heuristic}
\end{conjecture}
In fact in our experiments (where we apply method C of divide-and-select) this number is constant in most of the iterations. 

Consider the sub-routine $\mathcal{A}$. There are $N^2-1$ multiplications by $\chan{R(P_i)^{-1}}$ in each iteration for each unitary in $\widetilde{U}$. And by the above conjecture $|\widetilde{U}|\in \poly(N)$. Thus both the time and space complexity of $\mathcal{A}$ are in $\poly(N,m)$.

We have to call $\mathcal{A}$ at most $\tcount(U)$ times to solve MIN-T-SYNTH. Thus space and time complexity of MIN-T-SYNTH are in $\poly(N,\tcount(U))$.

\subsection{Implementations and results}

We implemented our heuristic algorithm 
MIN-T-SYNTH in Python on a machine with Intel(R) Core(TM) i7-7700K CPU at 4.2GHz, having 8 cores and 16 GB RAM and running Linux Debian 9.12. We have observed that in the first level of the tree for every iteration of MIN-T-SYNTH the sde of all the nodes increases for all the untiaries we tested. So we joined the first two iterations i.e. we took set of nodes with minimum cardinality (with necessary restrictions) in level 2 of the tree (considering root is at level 0). After that we perform the rest of the iterations successively, as described. This improved the running time. We also note here that we can synthesize part of the circuit (except the Clifford $C_0$) from the output of MIN-T-SYNTH using Fact \ref{fact:cliffConj} (Section \ref{subsec:decompose}). We can use the algorithm in \cite{2004_AG} to synthesize $C_0$. This procedure is very efficient. The main bottleneck is to get the decomposition.

\begin{table}
\centering
\footnotesize
 \begin{tabular}{|p{1cm}|p{0.5cm}|p{0.5cm}|p{0.8cm}|p{0.3cm}|p{1cm}|p{0.5cm}|}
  \hline
  Unitary & $\#$q & T-count & Time & $\#$c & Time \cite{2016_dMM} & $\#$c \cite{2016_dMM}   \\
  \hline
  Toffoli & 3 & 7 & 5.75s & 1 & 25.9870s & 4096 \\
  \hline 
  Fredkin & 3 & 7 & 5.9s & 1 & 25.0031s & 4096 \\
  \hline 
  Peres & 3 & 7 & 5.74s & 1 & 25.4931s & 4096 \\
  \hline
  Quantum OR & 3 & 7 & 5.74s & 1 & 24.1854s & 4096 \\
  \hline
  Negated Toffoli & 3 & 7 & 5.75s & 1 & 26.9162s & 4096 \\
  \hline 
  1-bit full adder & 4 & 7 & 429.17s & 1 & 12.5hr & 4096 \\
  \hline 
  $U_1$ & 4 & 11 & 2.17hr & 1 & N/A & N/A  \\
  \hline 
  $U_2$ & 4 & 7 & 391.27s & 1 & N/A & N/A \\
  \hline
 \end{tabular}
 \caption{Comparison of running time of MIN-T-SYNTH (serial) with the parallel algorithm in \cite{2016_dMM}. Time given in hours (hr), minutes (min) and seconds (s). $\#$q denotes the number of qubits and $\#c$ denotes the number of cores used.}
 \label{tab:runTime}
\end{table}
\begin{table}[h]
\centering
\footnotesize
 \begin{tabular}{|p{0.3cm}|p{0.7cm}|p{1cm}|p{1cm}|p{1.1cm}|p{1.1cm}|}
  \hline
 $\#$q & $\#$T & Max $\#$nodes (mean) & Max $\#$nodes (std) & Time (mean) & Time (std) \\
 \hline
 \multirow{4}{*}{$2$} & 10 & 7.3 & 1.273 & 0.0551s & 0.0094s \\
 \cline{2-6}
 & 20 & 45.2 & 17.89 & 1.1926s & 0.6587s \\
 \cline{2-6} 
 & 30 & 135.6 & 44.26 & 7.556s & 2.563s \\
 \cline{2-6}
 & 40 &  933.4 & 282.098 & 2663.568s & 99.8533s \\
 \hline
 \multirow{2}{*}{$3$} & 10 & 19.1 & 11.086 & 1.375s & 0.439s \\
 \cline{2-6}
 & 20 & 434.8 & 198.348 & 235.7675s & 118.73s \\
 \cline{2-6}
 & 30 & 1735 & 1056.78 & 1303.71s & 746.3s \\
 \hline
 \multirow{1}{*}{$4$} & 10 & 10.3 & 2.561 & 55.610s & 12.545s \\
 \hline
 \end{tabular}
 \caption{Performance of MIN-T-SYNTH on some random circuits. $\#q$ denotes the number of qubits and $\#T$ denotes the number of T-gates used to generate the input unitary.}
 \label{tab:random}
 \end{table}

We tested some known 3 and 4 qubit gates like Toffoli, Fredkin, Peres, Quantum OR, Negated Toffoli and also 4-qubit 1-bit reversible adder circuit. We found the T-count for all these gates is 7. This is in agreement with the results in \cite{2013_AMMR,2013_GKMR,2016_dMM} but our algorithm is much more efficient. For example the largest circuit that could be synthesized with the parallel algorithm in \cite{2016_dMM} was the 4-qubit 1-bit reversible adder and it took 12.5 hours using 4096 cores (processors). In comparison our MIN-T-SYNTH takes about 7 min 9 sec to get a decomposition into $R(P)$s using 1 core. It takes an additional few seconds to synthesize a circuit for this part.

We also tested some 4-qubit circuits like 
$U_1=(TOF \otimes \id)(\id \otimes TOF)$ \cite{2014_AMMR} and $U_2=(TOF \otimes \id)(\id \otimes TOF)(TOF \otimes \id)$, where $TOF$ is the 3-qubit Toffoli gate. We obtained a T-count of 11 and 7 respectively instead of 14 and 21, had we just plugged in an optimized circuit for Toffoli. The re-synthesis algorithm Tpar returned a circuit with 12 T gates for $U_1$ \cite{2014_AMMR}.A comparison of performance of our algorithm (serial) with the parallel algorithm in \cite{2016_dMM} has been given in Table \ref{tab:runTime}. (The running time of our algorithm does not include the synthesis of $C_0$, but this part is usually more efficient.) We have compared the performance of our serial algorithm with the fastest (parallel) synthesis algorithm we know.

\begin{figure}[h]
\centering
\includegraphics[width=16cm, height=1.75cm]{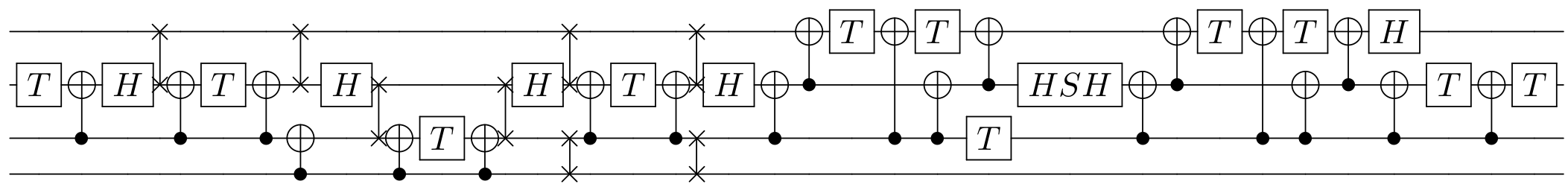}
\caption{A partial circuit of $U_1$ over the Clifford+T gate set. The circuit has been drawn from the first T to last T gate. The path obtained for $U_1$ was $[\id\id\Y\id, \id\Z\Y\id,\id\id\Y\X,\id\Z\Y\X,\id\id\X\X,\id\Z\id\id,\id\Z\X\X,\Z\id\id\X,$\\$\Z\id\X\id,\Z\Z\id\X,\Z\Z\X\id, \chan{C_0}]$ for some $C_0\in\cliff_n$.}
\label{fig:TT}
\end{figure}

\begin{figure}[h]
\centering
\includegraphics[width=10cm, height=1.75cm]{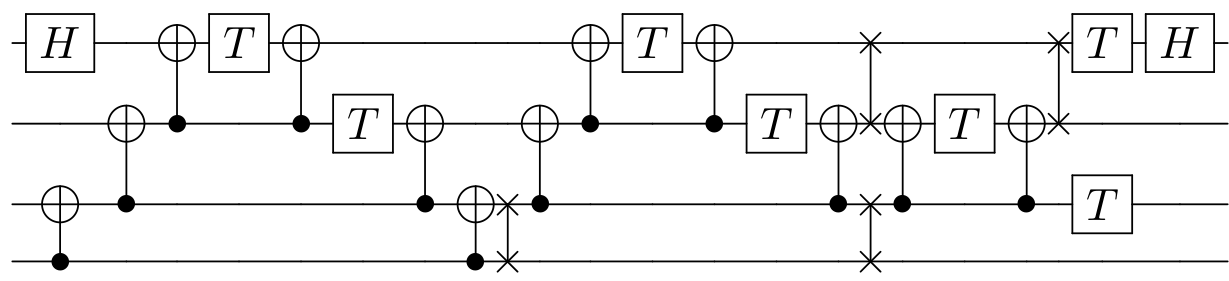}
\caption{A partial circuit of $U_2$ over the Clifford+T gate set. The circuit has been drawn from the first T to last T gate. The path obtained for $U_1$ was $[\id\id\id\X,\id\Z\id\id,\id\Z\id\X,\Z\id\Z\id,\Z\id\Z\X,\Z\Z\Z\id,\Z\Z\Z\X,\chan{C_0}]$ for some $C_0\in\cliff_n$.}
\label{fig:TTT}
\end{figure}

For completeness we have given the circuits for the known unitaries like Toffoli, Fredkin, Peres and 1-bit reversible adder in Figure \ref{app:fig:TFP3} and \ref{app:fig:adder} of Appendix \ref{app:circuits}. We drew a partial circuit (from the first T to last T gate) of $U_1$ and $U_2$ in Fig \ref{fig:TT} and \ref{fig:TTT} respectively. In the figures the connected $\X$ gates indicate swaps which can be implemented by CNOT gates.

In order to test the speed of our method as T-count increases we synthesized several random 2, 3 and 4 qubit unitaries formed using 10, 20, 30 and 40 T-gates (Table \ref{tab:random}). As mentioned in the introduction, 10, 20, 30 and 40 serve as upper bounds on the T-counts of the resulting unitaries. Using only unitaries as input our algorithm was able to synthesize all of them (return a decomposition) with T-count equal to or less than the upper bound. We synthesized 10 random circuits in each row of Table \ref{tab:random}. Of course, we cannot verify that we obtained T-count for all circuits because to the best of our knowledge, there are no existing algorithms that can be run on a PC to return the T-count of unitaries corresponding to such large circuits. A graph showing the distribution of the average number of unitaries (or nodes) selected in each iteration of MIN-T-SYNTH has been given in Appendix \ref{app:circuits} (Figure \ref{graph:Q}). We have also included the maximum number of unitaries (or nodes) accumulated (on an average) during any iteration of MIN-T-SYNTH. These are evidences to show that the complexity of our algorithm is much less than exponential in $N$, thus indicating that Conjecture \ref{conj:heuristic} may be correct for many unitaries.

\section{CONCLUSION}
\label{sec:conclude}

In this paper we have studied the complexity of synthesizing a T-count-optimal circuit for a given unitary. Modifying the meet-in-the-middle algorithm of \cite{2013_GKMR} we gave a provable nested meet-in-the-middle algorithm for COUNT-T (decision), where we are required to decide if an exactly synthesizable unitary $U$ can be implemented with at most $m$ T-gates. Solving this decision problem helps us to calculate the T-count as well as synthesize a T-count-optimal circuit with overhead polynomial in the dimension of $U$ and the T-count. Our algorithm has better space complexity but worse time complexity. However, both the time and space complexity is exponential in $\tcount(U)$, the T-count of $U$. 

We gave a heuristic algorithm for MIN-T-SYNTH, where we are required to synthesize a T-count-optimal circuit for $U$. This algorithm has space and time complexity polynomial in $N$ and $\tcount(U)$, thus (partly) answering a question left open in \cite{2013_GKMR}. 
While our method still scales exponentially with the number of qubits (though with a lower exponent), there is a large improvement by going from exponential to polynomial scaling with $\tcount(U)$. 
We tested our (serial) algorithm on some 3 and 4 qubit circuits and were able to obtain T-count with a much lower running time compared to previous algorithms like the parallel one in \cite{2016_dMM}. 



\section*{ACKNOWLEDGEMENT}
We thank Jo$\tilde{a}$o Marcos Vensi Basso for some useful discussions. We also thank Vlad Gheorghiu for allowing us to use his office computer on which some of the implementations were done. The authors would like to thank the anonymous reviewers for their helpful comments.


\section*{DATA AVAILABILITY}

The channel representation of the unitaries considered in this paper and the results obtained are available in a Github repository at \cite{data}.

\section*{CODE AVAILABILITY}
The software implementations and experiments were done by P. Mukhopadhyay.
The code is available from the authors on reasonable request.

\appendix
\section*{APPENDIX}

\section{Some additional preliminaries}
\label{app:prelim}

\subsection*{Cliffords and Paulis}
\label{prelim:cliffPauli}

The \emph{single qubit Pauli matrices} are as follows:
\begin{eqnarray}
 \X=\begin{bmatrix}
     0 & 1 \\
    1 & 0
    \end{bmatrix} \qquad  
 \Y=\begin{bmatrix}
     0 & -i \\
     i & 0
    \end{bmatrix} \qquad 
 \Z=\begin{bmatrix}
     1 & 0 \\
     0 & -1
    \end{bmatrix}
\label{eqn:Pauli1}
\end{eqnarray}
Parenthesized subscripts are used to indicate qubits on which an operator acts. For example, $\X_{(1)}=\X\otimes\id^{\otimes (n-1)}$ implies that Pauli $\X$ matrix acts on the first qubit and the remaining qubits are unchanged.

The \emph{$n$-qubit Pauli operators} are :
\begin{eqnarray}
 \pauli_n=\{Q_1\otimes Q_2\otimes\ldots\otimes Q_n:Q_i\in\{\id,\X,\Y,\Z\} \}.
 \label{eqn:paulin}
\end{eqnarray}

The \emph{single-qubit Clifford group} $\cliff_1$ is generated by the Hadamard and phase gate.
\begin{eqnarray}
 \cliff_1=\braket{\had,\phase} 
 \label{eqn:cliff1}
\end{eqnarray}
where
\begin{eqnarray}
 \had=\frac{1}{\sqrt{2}}\begin{bmatrix}
       1 & 1 \\
       1 & -1
      \end{bmatrix}\qquad 
 \phase=\begin{bmatrix}
       1 & 0 \\
       0 & i
      \end{bmatrix}.
\end{eqnarray}
When $n>1$ the \emph{$n$-qubit Clifford group} $\cliff_n$ is generated by these two gates (acting on any of the $n$ qubits) along with the two-qubit $\CNOT=\ket{0}\bra{0}\otimes\id+\ket{1}\bra{1}\otimes\X$ gate (acting on any pair of qubits). 

Cliffords map Paulis to Paulis up to a possible phase of $-1$ i.e. for any $P\in\pauli_n$ and any $C\in\cliff_n$ we have
$$
    CPC^{\dagger}=(-1)^bP'
$$
for some $b\in\{0,1\}$ and $P'\in\pauli_n$. In fact, given two Paulis (neither equal to the identity), it is always possible to efficiently find a Clifford which maps one to the other.
\begin{fact}[\cite{2013_GKMR}]
 For any $P,P'\in\pauli_n\setminus\{\id\} $ there exists a Clifford $C\in\cliff_n$ such that $CPC^{\dagger}=P'$. A circuit for $C$ over the gate set $\{\had,\phase,\CNOT\}$ can be computed efficiently (as a function of $n$).
 \label{fact:cliffConj}
\end{fact}

\subsection*{The group $\mathcal{J}_n$ generated by Clifford and $\T$ gates}

The group $\mathcal{J}_n$ is generated by the $n$-qubit Clifford group along with the $\T$ gate, where
\begin{eqnarray}
 \T=\begin{bmatrix}
     1 & 0 \\
     0 & e^{i\frac{\pi}{4}}
    \end{bmatrix}
\end{eqnarray}

Thus for a single qubit
$
 \mathcal{J}_1 = \braket{\had,\T}    
$
and for $n>1$ qubits
$
 \mathcal{J}_n=\braket{\had_{(i)},\T_{(i)},\CNOT_{(i,j)}:i,j\in [n]}.
$
It can be easily verified that $\mathcal{J}_n$ is a group, since the $\had$ and $\CNOT$ gates are their own inverses and $\T^{-1}=\T^7$. Here we note $\phase=\T^2$.

We denote the group of unitaries exactly synthesized over the Clifford+T gate set by $\clifft_n$. Some elements of this group cannot be exactly synthesized over this gate set without ancilla qubits. 

The following characterization of this group was proved by Giles and Selinger \cite{2013_GS}.

\begin{theorem} (Theorem 1 and Corollary 2 from \cite{2013_GS})
Let $U(N)$ (where $N=2^n$) is the group of $n$-qubit unitaries. Then the following are equivalent.
\begin{enumerate}
 \item $U$ can be exactly represented by a quantum circuit over the Clifford+T gate set, possibly using some finite number of ancillas that are initialized and finalized in state $\ket{0}$.
 
 If ancilla is required, then a single ancilla is sufficient.
 
 \item The entries of $U$ belong to the ring $\intg\left[i,\frac{1}{\sqrt{2}}\right]$.
where 
 \begin{eqnarray}
 \intg\left[i,\frac{1}{\sqrt{2}}\right]&=&\Big\{(a+bi+c\sqrt{2}+di\sqrt{2})/\sqrt{2}^k  
 :a,b,c,d\in\intg,\quad k\in\nat \Big\}.  \nonumber
\end{eqnarray}
 
 \item No ancilla is required if $\det(U)=e^{i\frac{\pi}{8}Nr} $ for some $r\in[8]$. For $n\geq 4$ the condition on the determinant is simply $\det(U)=1$.
\end{enumerate}
 \label{thm:2013_GS}
\end{theorem}

\subsection*{Proof of some results in Section \ref{sec:prelim}}

\begin{lemma}
Let $V$ and $U$ are $N_1\times N_1$ and $N_2\times N_2$ unitaries respectively, where $N_1=2^{m}$ and $N_2=2^{n}$. Then
$
    \chan{(V\otimes U)}=\chan{V}\otimes\chan{U}.
$
 \label{app:lem:chanAncilla2}
\end{lemma}
\begin{proof}
 $\chan{(V\otimes U)}$ is a $N_1^2N_2^2\times N_1^2N_2^2$ matrix. Any index $(r,s)$ can be written as $r=N_2^2(r_1-1)+r_2$ and $s=N_2^2(s_1-1)+s_2$. Let $P_r=\bigotimes_{i=1}^m B_i\bigotimes_{j=m+1}^{m+n}B_j=\widetilde{B_1}\otimes \widetilde{B_2}$ and $P_s=\bigotimes_{i=1}^m C_i\bigotimes_{j=m+1}^{m+n}C_j=\widetilde{C_1}\otimes \widetilde{C_2}$ are the row and column indices respectively. Here $\widetilde{B_1}=\bigotimes_{i=1}^m B_i, \widetilde{B_2}=\bigotimes_{j=m+1}^{m+n} B_j, \widetilde{C_1}=\bigotimes_{i=1}^m C_i$ and $\widetilde{C_2}=\bigotimes_{j=m+1}^{m+n} C_j$. Think of these as base 4 strings.
 \begin{eqnarray}
  (\chan{V\otimes U})_{rs} 
  &=& \frac{1}{2^{m+n}} \tr\left[(\widetilde{B_1}\otimes\widetilde{B_2})(V\otimes U)(\widetilde{C_1}\otimes\widetilde{C_2})\right]   \nonumber \\
  &=&\frac{1}{2^m}\tr(\widetilde{B_1}V\widetilde{C_1})\frac{1}{2^n}\tr(\widetilde{B_2}U\widetilde{C_2})   \nonumber \\
  &=&\chan{V}_{r_1s_1}\chan{U}_{r_2s_2} 
  =(\chan{V}\otimes\chan{U})_{N_2^2(r_1-1)+r_2,N_2^2(s_1-1)+s_2}  \nonumber
 \end{eqnarray}
This proves the lemma.
\end{proof}

\section{Proof of properties of $\chan{R(P)}$}
\label{app:RP}

We give the proof of Claim \ref{claim:diagOffdiag} in Section \ref{sec:chanRep_n}.
\begin{claim}
\begin{enumerate}
    \item If a diagonal entry is $1$ then all other entries in the corresponding row and column is $0$.
    \item If a diagonal entry is $\frac{1}{\sqrt{2}}$ then one other entry in the corresponding row is $\pm \frac{1}{\sqrt{2}}$ and one other entry in the corresponding column is $\mp\frac{1}{\sqrt{2}}$.
 \end{enumerate}
  \label{app:claim:diagOffdiag}
\end{claim}
\begin{proof}
1. From Claim \ref{claim:diagonal} we note that a diagonal entry is $1$ if for even number of (co-ordinates) $j$ we have $B_jA_j=\pm iG_j$ where $G_j\in\pauli_1\setminus\{\id\}$ and for the rest we have $B_jA_j=G_j'$ where $G_j'\in\pauli_1$. In the former case we have $A_j,B_j\in\pauli_1\setminus\{\id\}$ and $A_j\neq B_j$ and in the latter at least one is identity. W.l.o.g. let $1,\ldots,m'$ are the places where the former case holds. Then in the remaining row entries we can have $B_jC_j=\pm iA_j$ for even number of places or there exists $k\in[n]$ such that $B_jC_j\neq\pm A_j$ or $B_jC_jA_j\neq\id$. Thus by Claim \ref{claim:offDiagonal} these entries are $0$. Similar argument holds for the remaining entries in the column.

2. A diagonal entry is $\frac{1}{\sqrt{2}}$ if for an odd number of $j$ we have $B_jA_j=\pm iG_j$ and for the rest we have $B_jA_j=G_j'$ (Claim \ref{claim:diagonal}) where $G_j$ and $G_j'$ are as defined above. Note $G_j'=A_j$ or $B_j$ and at least one of $A_j$ or $B_j$ is $\id$ in the second case. Let w.l.o.g $1,\ldots,m$ are the co-ordinates where $B_jA_j=\pm iG_j$. Now consider the off-diagonal entries in the same row. It is not difficult to see that there can be only one column ($P_s=\bigotimes_{i=1}^nC_i$) in this row such that $B_jC_j=\pm iA_j$ for each $j\in[m]$ and $B_jC_jA_j=\id$ for $j\in[n]\setminus[m]$.
By Claim \ref{claim:offDiagonal} at this entry in the row we have $\pm\frac{1}{\sqrt{2}}$.

Similar argument holds for the column. Note for the column we have $C_j\leftarrow B_j$ and $B_j\leftarrow C_j$, i.e. the indices get swapped. So from Equation \ref{eqn:K'offDiag} we have the corresponding entry as $\mp\frac{1}{\sqrt{2}}$.

\end{proof}

Now we give the proof of Claim \ref{claim:diagSign} in Section \ref{sec:chanRep_n}.

\begin{claim}
 Exactly $2^{2n-1}$ diagonal elements can be $\frac{1}{\sqrt{2}}$.
 \label{app:claim:diagSign}
 \end{claim}
 \begin{proof}
 It is enough to calculate the number of $\bigotimes_{j=1}^n B_j$ that can satisfy the conditions of Claim \ref{claim:diagonal}. We must have $B_jA_j=\pm i G_j$ (where $G_j\in\pauli_1\setminus\{\id\}$) at odd number of $j\in[n]$. Equivalently at these co-ordinates $A_j\neq B_j$ where $A_j,B_j\in\pauli_1\setminus\{\id\}$. 
 
 Let $\bigotimes_{j=1}^n A_j$ has non-identity Paulis in $m$ co-ordinates. Without loss of generality let $m$ is odd and the co-ordinates are $[m]$. Given $A_j$ there can be two possible values of $B_j$ to satisfy $B_jA_j= \pm i G_j$. In the remaining places if $A_j$ is non-identity then by condition of Claim \ref{claim:diagonal} $B_j=A_j$ or $B_j=\id$.  Thus number of $j\in[m]$ satisfying this condition is 
 \begin{eqnarray}
    \binom{m}{1}22^{m-1}+\binom{m}{3}2^32^{m-3}+\cdots+\binom{m}{m}2^m2^0  
    = 2^{2m-1}   \nonumber
 \end{eqnarray}
 Now in the co-ordinates $j\in\{m+1,\ldots,n\}$ we must have $(B_jA_j)^2=\id$. Since $A_j=\id$ so $B_j$ can be any of the Paulis. Thus total number of ways we can satisfy the conditions of Claim \ref{claim:diagonal} is
 $$
    2^{2m-1}4^{n-m}=2^{2m-1+2n-2m}=2^{2n-1}
 $$
 Thus there can be exactly $2^{2n-1}$ diagonal elements with $\frac{1}{\sqrt{2}}$.
 \end{proof}

\begin{figure}[h]
\centering
\includegraphics[width=8cm, height=7.5cm]{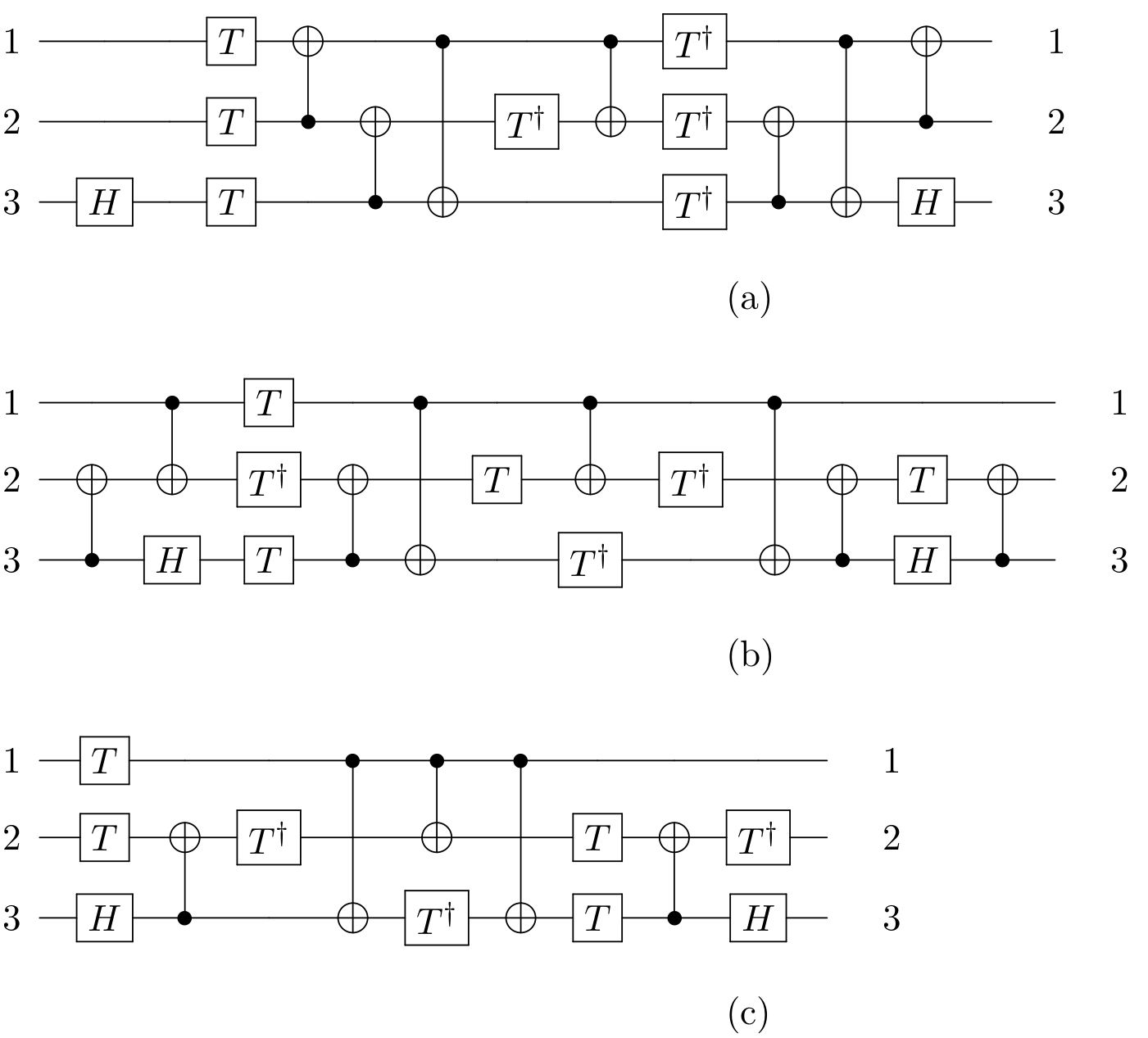}
\caption{Decomposition of the (a) Toffoli gate, (b) Fredkin gate and (c) Peres gate over the Clifford+T gate set.}
\label{app:fig:TFP3}
\end{figure}

\begin{figure}[h]
\centering
\includegraphics[width=10cm, height=2cm]{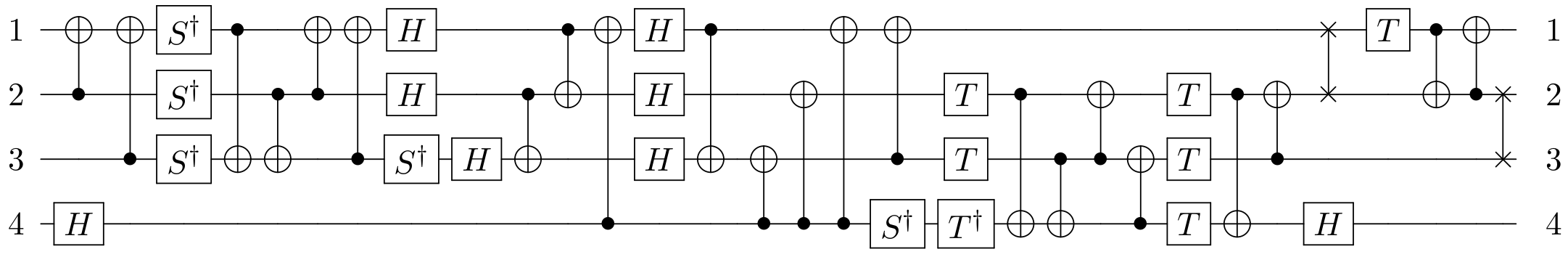}
\caption{Decomposition of 4-qubit adder over the Clifford+T gate set. }
\label{app:fig:adder}
\end{figure}

\begin{figure}
 \centering
 \begin{subfigure}[b]{0.475\textwidth}
 \centering
  \includegraphics[width=7cm,height=5cm]{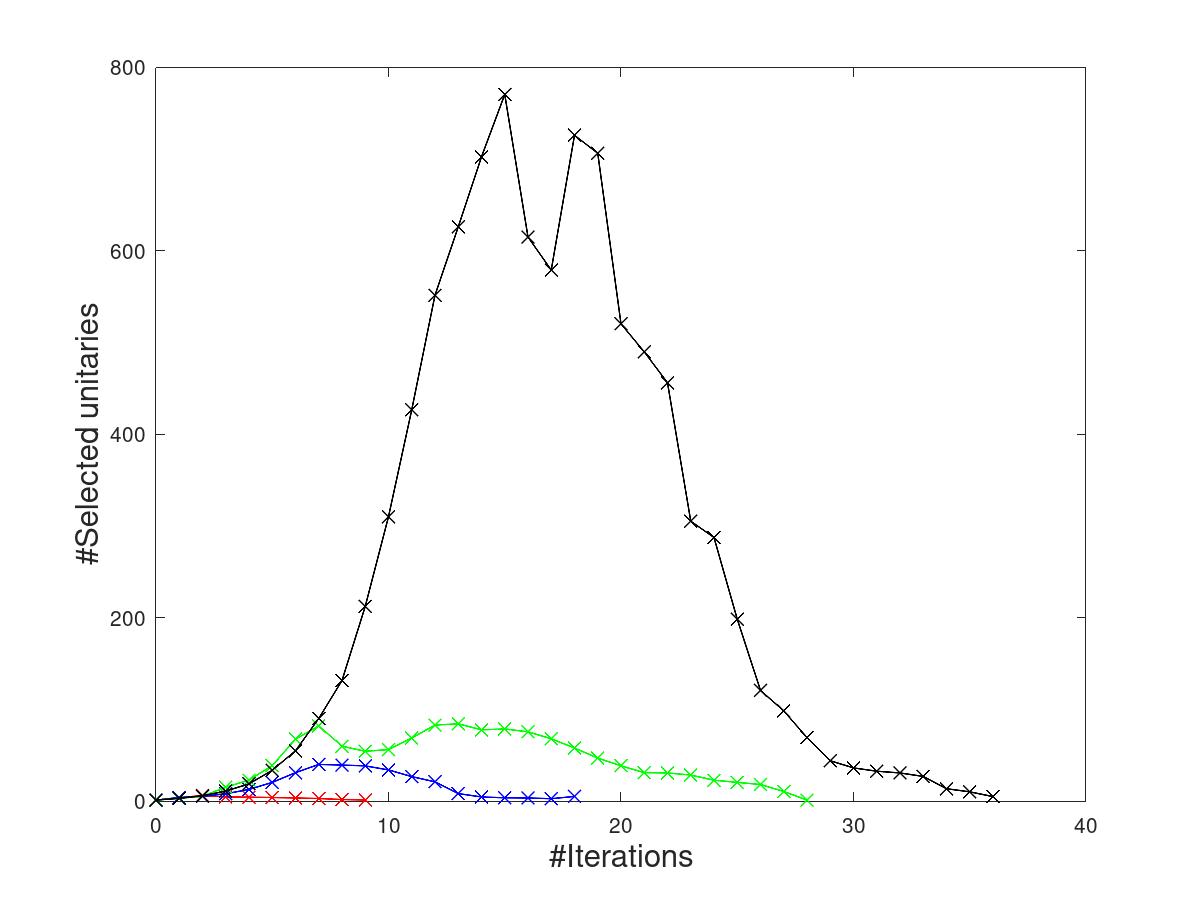}
  \end{subfigure}
 \hfill
 \begin{subfigure}[b]{0.475\textwidth}
 \centering
 \includegraphics[width=7cm,height=5cm]{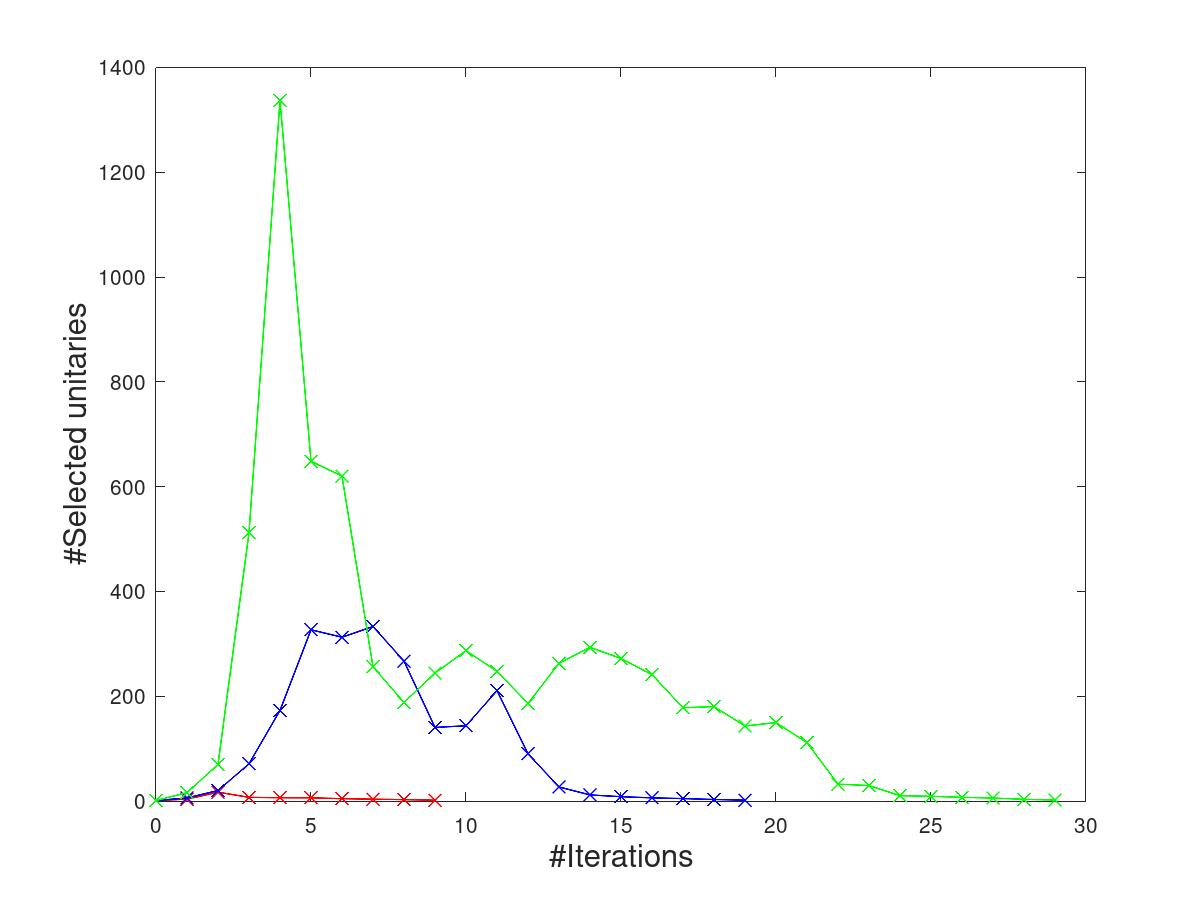}
 \end{subfigure}
 \vskip\baselineskip
 \begin{subfigure}[b]{0.475\textwidth}
 \centering
 \includegraphics[width=7cm,height=5cm]{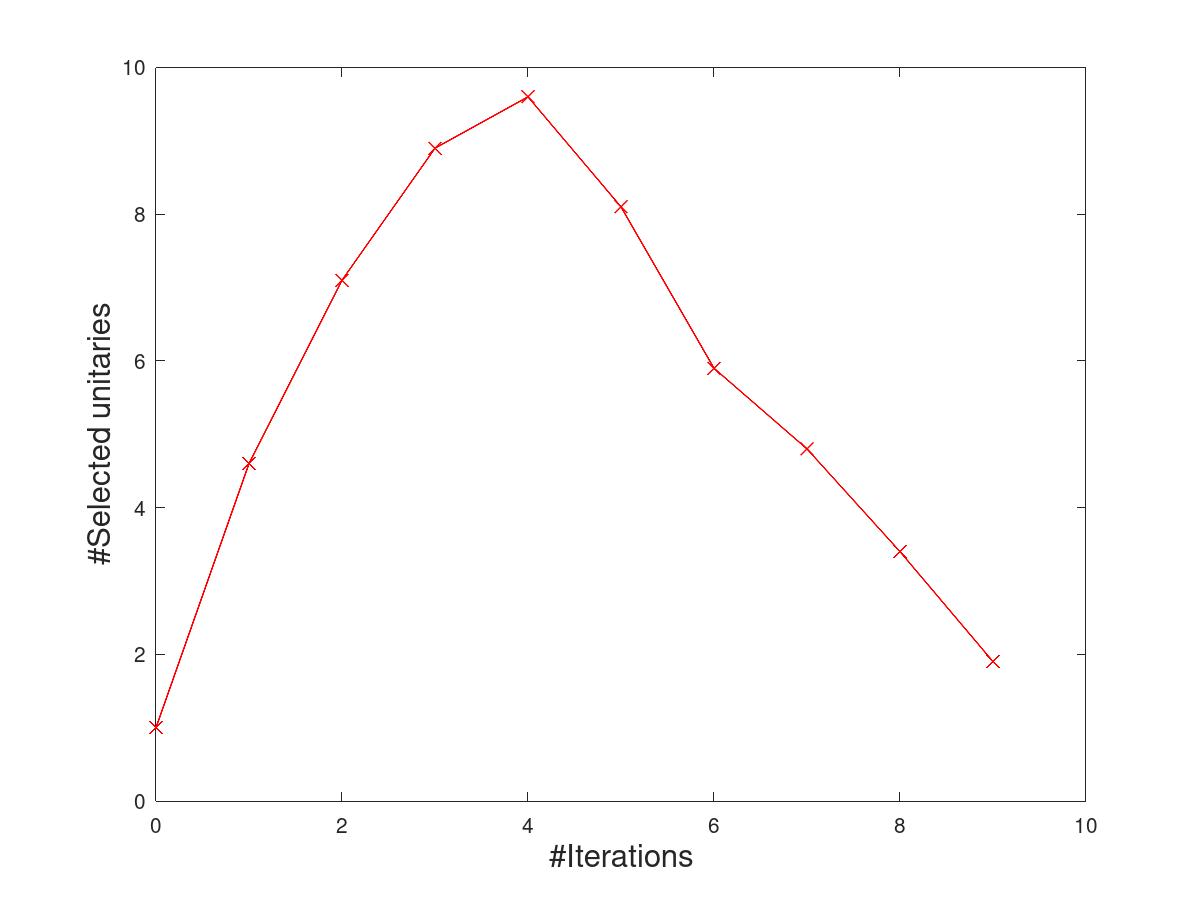}
  \end{subfigure}
 \hfill 
  \caption{The average number of untiaries selected in each iteration during the implementation of 2 qubit (leftmost plot), 3 qubit (centre plot) and 4 qubit (rightmost plot) unitaries. The red, blue, green and black curves are for unitaries obtained from circuits with at most 10, 20, 30 and 40 T gates respectively.}
\label{graph:Q}
\end{figure}

\section{Proof of Theorem \ref{thm:nestMeet}}
\label{app:provable}

In this section we give the proof of Theorem \ref{thm:nestMeet} in Section \ref{sec:algo}.

\begin{theorem}
Let $m\in\nat$, $c\geq 2$ and $j\geq 1$ are integers. $U\in\clifft_n$ is a unitary with $j\lceil\frac{m}{c}\rceil<\tcount(U)\leq B$, where $B=\min\{(j+1)\lceil\frac{m}{c}\rceil,m\}$. Let $\database_0^n,\database_1^n,\ldots,\database_{\lceil\frac{m}{c}\rceil}^n$ are sorted coset databases.

Then $t=\tcount(U)$ is the smallest integer in $\{j\lceil\frac{m}{c}\rceil+1,j\lceil\frac{m}{c}\rceil+2,\ldots,B\}$ for which
\begin{eqnarray}
 \left(W^{\dag}\chan{U} \right)=V
 \label{eqn:thmNestMeet}
\end{eqnarray}
with $W^{(co)}\in\database_r^n$, where $r=B-j\lceil\frac{m}{c}\rceil$. $V$ is a unitary such that $\tcount(V)\leq j\lceil\frac{m}{c}\rceil$.
 \label{app:thm:nestMeet}
\end{theorem}

\begin{proof}
 Using Theorem \ref{thm:cosetLabel} we see that Equation \ref{eqn:thmNestMeet} implies
 $
  \chan{U}=WVC  
 $
for some $C\in\chan{\cliff_n}$. Hence, whenever Equation \ref{eqn:thmNestMeet} holds we have
\begin{eqnarray}
 \tcount(U)&\leq&\tcount(W)+\tcount(V)=r+\tcount(V) 
 =B-j\left\lceil\frac{m}{c}\right\rceil+\tcount(V)    \nonumber
\end{eqnarray}
Now since the algorithm reached the $j^{th}$ iteration it implies that $\tcount(U)>j\lceil\frac{m}{c}\rceil$ and we are checking if it is less than $B$. Also $\tcount(W)= r$, so $0\leq\tcount(V)\leq j\lceil\frac{m}{c}\rceil$ and since the property holds till the $(j-1)^{th}$ iteration so the recursive step correctly returns it and we get $\tcount(U)$ as the above sum.

For completion we prove that Equation \ref{eqn:thmNestMeet} holds with $t=\tcount(U)$. From Theorem \ref{thm:decompose} we can write 
$
    \chan{U}=W_0V_0C_0 
$
where $C_0\in\chan{\cliff_n}$ and
$$
    W_0=\prod_{i=B}^{j\lceil\frac{m}{c}\rceil+1} \chan{R(P_i)} \quad \text{ and }\quad
    V_0=\prod_{i=t+j\lceil\frac{m}{c}\rceil-B}^{1} \chan{R(P_i)}
$$
for some Paulis $P_i\in\pauli_n\setminus\{\id\}$. Note $\tcount(W_0)=B-j\left\lceil\frac{m}{c}\right\rceil=r$ and $\tcount(V_0)=t+j\left\lceil\frac{m}{c}\right\rceil-B$.

From property 3 of Definition \ref{defn:sortCoset} there exists $W\in\database_r^n$ satisfying $W^{(co)}=W_0^{(co)}$, which implies
$
    WC_1=W_0
$
for some $C_1\in\chan{\cliff_n}$ by Theorem \ref{thm:decompose}. Hence
$
    \chan{U}=WC_1V_0C_0.
$
Now 
$\tcount(C_1V_0C_0)=\tcount(V_0)=j\left\lceil\frac{m}{c}\right\rceil-(B-t)\leq j\left\lceil\frac{m}{c}\right\rceil$. By our assumption the algorithm correctly returns $\tcount(C_1V_0C_0)$ by a recursive loop. This implies there exists some $V$ and some $C_2\in\chan{\cliff_n}$ such that $\tcount(VC_2)=\tcount(C_1V_0C_0)$ and their coset labels are equivalent. Thus
$
    \chan{U}=WVC_2
$
or equivalently $W^{\dag}\chan{U}=VC$. Applying Theorem \ref{thm:decompose} we get $\left(W^{\dag}\chan{U}\right)^{(co)}=V^{(co)}$.
\end{proof}

\section{Some quantum circuits and graphs}
\label{app:circuits}

In this section first we give the complete circuits (Figure \ref{app:fig:TFP3} and \ref{app:fig:adder}) of some untiaries implemented by our algorithm. We have given the circuits known in literature like in \cite{2013_GKMR} and \cite{2016_dMM}.

Then we give graphs that show the distribution of average number of unitaries selected during each iteration of algorithm MIN-T-SYNTH. The average is over the number of 2, 3 and 4 qubit unitaries (Figure \ref{graph:Q}) implemented by us. In the graphs the red, blue, green and black curves show the distribution for unitaries obtained from circuits with at most 10, 20, 30 and 40 T gates respectively. We find that on an average the number of unitaries first increases and then decreases. The maximum number of unitaries is still polynomial in the size of the unitary.

\newcommand{\etalchar}[1]{$^{#1}$}

\begin{thebibliography}{AMMR13}

\bibitem{1994_S}
Peter~W Shor.
\newblock Algorithms for quantum computation: Discrete logarithms and
  factoring.
\newblock In {\em Proceedings 35th annual symposium on foundations of computer
  science}, pages 124--134. Ieee, 1994.

\bibitem{1999_S}
Peter~W Shor.
\newblock Polynomial-time algorithms for prime factorization and discrete
  logarithms on a quantum computer.
\newblock {\em SIAM review}, 41(2):303--332, 1999.

\bibitem{1998_G}
Daniel Gottesman.
\newblock The heisenberg representation of quantum computers.
\newblock {\em arXiv preprint quant-ph/9807006}, 1998.

\bibitem{2004_AG}
Scott Aaronson and Daniel Gottesman.
\newblock Improved simulation of stabilizer circuits.
\newblock {\em Physical Review A}, 70(5):052328, 2004.

\bibitem{2005_BK}
{\sc Bravyi, S., and Kitaev, A.}
\newblock Universal quantum computation with ideal clifford gates and noisy
  ancillas.
\newblock {\em Physical Review A 71}, 2 (2005), 022316.

\bibitem{1999_GC}
Daniel Gottesman and Isaac~L Chuang.
\newblock Quantum teleportation is a universal computational primitive.
\newblock {\em arXiv preprint quant-ph/9908010}, 1999.

\bibitem{2009_FSG}
Austin~G Fowler, Ashley~M Stephens, and Peter Groszkowski.
\newblock High-threshold universal quantum computation on the surface code.
\newblock {\em Physical Review A}, 80(5):052312, 2009.

\bibitem{2006_AGP}
Panos Aliferis, Daniel Gottesman, and John Preskill.
\newblock Quantum accuracy threshold for concatenated distance-3 codes.
\newblock {\em Quantum Information \& Computation}, 6(2):97--165, 2006.

\bibitem{2018_G}
Craig Gidney.
\newblock Halving the cost of quantum addition.
\newblock {\em Quantum}, 2:74, 2018.

\bibitem{2016_BG}
Sergey Bravyi and David Gosset.
\newblock Improved classical simulation of quantum circuits dominated by
  clifford gates.
\newblock {\em Physical review letters}, 116(25):250501, 2016.

\bibitem{2016_BSS}
Sergey Bravyi, Graeme Smith, and John~A Smolin.
\newblock Trading classical and quantum computational resources.
\newblock {\em Physical Review X}, 6(2):021043, 2016.

\bibitem{2013_PR}
Adam Paetznick and Ben~W Reichardt.
\newblock Universal fault-tolerant quantum computation with only transversal
  gates and error correction.
\newblock {\em Physical review letters}, 111(9):090505, 2013.

\bibitem{2003_K}
A~Yu Kitaev.
\newblock Fault-tolerant quantum computation by anyons.
\newblock {\em Annals of Physics}, 303(1):2--30, 2003.

\bibitem{2012_BSK_}
Joseph~W Britton, Brian~C Sawyer, Adam~C Keith, C-C~Joseph Wang, James~K
  Freericks, Hermann Uys, Michael~J Biercuk, and John~J Bollinger.
\newblock Engineered two-dimensional ising interactions in a trapped-ion
  quantum simulator with hundreds of spins.
\newblock {\em Nature}, 484(7395):489, 2012.

\bibitem{2011_BWC_}
Kenton~R Brown, Andrew~C Wilson, Yves Colombe, C~Ospelkaus, Adam~M Meier,
  E~Knill, D~Leibfried, and David~J Wineland.
\newblock Single-qubit-gate error below 10- 4 in a trapped ion.
\newblock {\em Physical Review A}, 84(3):030303, 2011.

\bibitem{2012_CGC_}
Jerry~M Chow, Jay~M Gambetta, AD~C{\'o}rcoles, Seth~T Merkel, John~A Smolin,
  Chad Rigetti, S~Poletto, George~A Keefe, Mary~B Rothwell, JR~Rozen, et~al.
\newblock Universal quantum gate set approaching fault-tolerant thresholds with
  superconducting qubits.
\newblock {\em Physical review letters}, 109(6):060501, 2012.

\bibitem{2012_RGP_}
Chad Rigetti, Jay~M Gambetta, Stefano Poletto, BLT Plourde, Jerry~M Chow,
  AD~C{\'o}rcoles, John~A Smolin, Seth~T Merkel, JR~Rozen, George~A Keefe,
  et~al.
\newblock Superconducting qubit in a waveguide cavity with a coherence time
  approaching 0.1 ms.
\newblock {\em Physical Review B}, 86(10):100506, 2012.

\bibitem{2012_BAOKM}
H{\'e}ctor Bombin, Ruben~S Andrist, Masayuki Ohzeki, Helmut~G Katzgraber, and
  Miguel~A Mart{\'\i}n-Delgado.
\newblock Strong resilience of topological codes to depolarization.
\newblock {\em Physical Review X}, 2(2):021004, 2012.

\bibitem{2012_FWH}
Austin~G Fowler, Adam~C Whiteside, and Lloyd~CL Hollenberg.
\newblock Towards practical classical processing for the surface code.
\newblock {\em Physical review letters}, 108(18):180501, 2012.

\bibitem{2020_MP}
Michele Mosca, and Marco Piani.
{\em Quantum threat timeline report 2020}.
\newblock \url{https://globalriskinstitute.org/download/quantum-threat-timeline-report-2020/}.

\bibitem{2008_PMH}
Ketan N Patel, Igor L Markov, and John P Hayes.
\newblock Optimal synthesis of linear reversible circuits.
\newblock{\em Quantum Inf. Comput.}, 8(3):282--294, 2008.

\bibitem{2005_DN}
CM~Dawson and MA~Nielsen.
\newblock The solovay-kitaev algorithm.
\newblock {\em Quantum Information and Computation}, 6(1):81--95, 2006.

\bibitem{2013_GS}
Brett Giles and Peter Selinger.
\newblock Exact synthesis of multiqubit clifford+ t circuits.
\newblock {\em Physical Review A}, 87(3):032332, 2013.

\bibitem{2013_K}
Vadym Kliuchnikov.
\newblock Synthesis of unitaries with clifford+ t circuits.
\newblock {\em arXiv preprint arXiv:1306.3200}, 2013.

\bibitem{2013_GKMR}
David Gosset, Vadym Kliuchnikov, Michele Mosca, and Vincent Russo.
\newblock An algorithm for the t-count.
\newblock {\em Quantum Information and Computation}, 14(15-16):1261--1276, 2014.

\bibitem{2016_dMM}
Olivia Di~Matteo and Michele Mosca.
\newblock Parallelizing quantum circuit synthesis.
\newblock {\em Quantum Science and Technology}, 1(1):015003, 2016.

\bibitem{2014_AMMR}
Matthew Amy, Dmitri Maslov, and Michele Mosca.
\newblock Polynomial-time t-depth optimization of clifford+ t circuits via
  matroid partitioning.
\newblock {\em IEEE Transactions on Computer-Aided Design of Integrated
  Circuits and Systems}, 33(10):1476--1489, 2014.

\bibitem{2018_HC}
Luke~E Heyfron and Earl~T Campbell.
\newblock An efficient quantum compiler that reduces t count.
\newblock {\em Quantum Science and Technology}, 4(1):015004, 2018.

\bibitem{2018_AAM}
Matthew Amy, Parsiad Azimzadeh, and Michele Mosca.
\newblock On the controlled-not complexity of controlled-not--phase circuits.
\newblock {\em Quantum Science and Technology}, 4(1):015002, 2018.

\bibitem{1997_K}
A~Yu Kitaev.
\newblock Quantum computations: algorithms and error correction.
\newblock {\em Russian Mathematical Surveys}, 52(6):1191, 1997.

\bibitem{2018_MT}
Edgard Mu{\~n}oz-Coreas and Himanshu Thapliyal.
\newblock Quantum circuit design of a t-count optimized integer multiplier.
\newblock {\em IEEE Transactions on Computers}, 68(5):729--739, 2018.

\bibitem{2021_TMK}
Himanshu Thapliyal, Edgard Mu{\~n}oz-Coreas, and Vladislav Khalus.
\newblock Quantum circuit designs of carry lookahead adder optimized for
  t-count t-depth and qubits.
\newblock {\em Sustainable Computing: Informatics and Systems}, 29:100457,
  2021.
  
\bibitem{2021_GKDetal}
SS~Gayathri, R~Kumar, Samiappan Dhanalakshmi, Gerard Dooly, and Dinesh~Babu
  Duraibabu.
\newblock T-count optimized quantum circuit designs for single-precision
  floating-point division.
\newblock {\em Electronics}, 10(6):703, 2021.

\bibitem{2013_KMM}
Vadym Kliuchnikov, Dmitri Maslov, and Michele Mosca.
\newblock Fast and efficient exact synthesis of single-qubit unitaries
  generated by clifford and t gates.
\newblock {\em Quantum Information \& Computation}, 13(7-8):607--630, 2013.

\bibitem{2016_RS}
Neil~J Ross and Peter Selinger.
\newblock Optimal ancilla-free clifford+ t approximation of z-rotations.
\newblock {\em Quantum Information \& Computation}, 16(11-12):901--953, 2016.

\bibitem{2013_J}
Cody Jones.
\newblock Low-overhead constructions for the fault-tolerant toffoli gate.
\newblock {\em Physical Review A}, 87(2):022328, 2013.

\bibitem{2014_lG}
Fran{\c{c}}ois Le~Gall.
\newblock Powers of tensors and fast matrix multiplication.
\newblock In {\em Proceedings of the 39th international symposium on symbolic
  and algebraic computation}, pages 296--303. ACM, 2014.

\bibitem{2013_AMMR}
Matthew Amy, Dmitri Maslov, Michele Mosca, and Martin Roetteler.
\newblock A meet-in-the-middle algorithm for fast synthesis of depth-optimal
  quantum circuits.
\newblock {\em IEEE Transactions on Computer-Aided Design of Integrated
  Circuits and Systems}, 32(6):818--830, 2013.

\bibitem{2013_S}
{\sc Selinger, P.}
\newblock Quantum circuits of t-depth one.
\newblock {\em Physical Review A 87}, 4 (2013), 042302.

\bibitem{2021_GMM}
Vlad Gheorghiu, Michele Mosca, and Priyanka Mukhopadhyay.
\newblock A quasi-polynomial time heuristic algorithm for synthesizing t-depth
  optimal circuits.
\newblock {\em arXiv preprint arXiv:2101.03142}, 2021.

\bibitem{2019_AM}
Matthew Amy and Michele Mosca.
\newblock T-count optimization and reed-muller codes.
\newblock {\em IEEE Transactions on Information Theory}, 2019.


\bibitem{2012_KMM}
Vadym Kliuchnikov, Dmitri Maslov, and Michele Mosca.
\newblock Fast and efficient exact synthesis of single qubit unitaries
  generated by clifford and t gates.
\newblock {\em Quantum Information \& Computation}, 13(7-8):607--630, 2013.

\bibitem{2013_KMM2}
Vadym Kliuchnikov, Dmitri Maslov, and Michele Mosca.
\newblock Asymptotically optimal approximation of single qubit unitaries by
  clifford and t circuits using a constant number of ancillary qubits.
\newblock {\em Physical review letters}, 110(19):190502, 2013.

\bibitem{2015_S}
Peter Selinger.
\newblock Efficient clifford+ t approximation of single-qubit operators.
\newblock {\em Quantum Information \& Computation}, 15(1-2):159--180, 2015.


\bibitem{2014_PS}
Adam Paetznick and Krysta~M Svore.
\newblock Repeat-until-success: non-deterministic decomposition of single-qubit
  unitaries.
\newblock {\em Quantum Information \& Computation}, 14(15-16):1277--1301, 2014.


\bibitem{2015_BRS}
Alex Bocharov, Martin Roetteler, and Krysta~M Svore.
\newblock Efficient synthesis of universal repeat-until-success quantum
  circuits.
\newblock {\em Physical review letters}, 114(8):080502, 2015.

\bibitem{2020_GRT}
Andrew N Glaudell, Neil J Ross, and Jacob M Taylor.
\newblock Optimal two-qubit circuits for universal fault-tolerant quantum computation.
\newblock {\em arXiv preprint arXiv:2001.05997}, 2020.

\bibitem{2019_dBBW}
Niel de~Beaudrap, Xiaoning Bian, and Quanlong Wang.
\newblock Techniques to reduce $\pi/4$-parity phase circuits, motivated by the zx calculus.
\newblock {\em arXiv preprint arXiv:1911.09039}, 2019.

\bibitem{2019_ZC}
Fang Zhang and Jianxin Chen.
\newblock Optimizing t gates in clifford+ t circuit as $\pi/4$ rotations around
  paulis.
\newblock {\em arXiv preprint arXiv:1903.12456}, 2019.

\bibitem{2019_KdW}
Aleks Kissinger and John van~de Wetering.
\newblock Reducing t-count with the zx-calculus.
\newblock {\em arXiv preprint arXiv:1903.10477}, 2019.

\bibitem{2014_ASD}
Nabila Abdessaied, Mathias Soeken, and Rolf Drechsler.
\newblock Quantum circuit optimization by hadamard gate reduction.
\newblock In {\em International Conference on Reversible Computation}, pages
  149--162. Springer, 2014.


\bibitem{data}
\url{https://github.com/PriyankaMukhopadhyay/Polynomial-time-T-count-algo}


\end{thebibliography}

\begin{thebibliography}{10}

\bibitem{2004_AG}
{\sc Aaronson, S., and Gottesman, D.}
\newblock Improved simulation of stabilizer circuits.
\newblock {\em Physical Review A 70}, 5 (2004), 052328.

\bibitem{2014_ASD}
{\sc Abdessaied, N., Soeken, M., and Drechsler, R.}
\newblock Quantum circuit optimization by hadamard gate reduction.
\newblock In {\em International Conference on Reversible Computation\/} (2014),
  Springer, pp.~149--162.

\bibitem{2006_AGP}
{\sc Aliferis, P., Gottesman, D., and Preskill, J.}
\newblock Quantum accuracy threshold for concatenated distance-3 codes.
\newblock {\em Quantum Information \& Computation 6}, 2 (2006), 97--165.

\bibitem{2014_AMMR}
{\sc Amy, M., Maslov, D., and Mosca, M.}
\newblock Polynomial-time t-depth optimization of clifford+ t circuits via
  matroid partitioning.
\newblock {\em IEEE Transactions on Computer-Aided Design of Integrated
  Circuits and Systems 33}, 10 (2014), 1476--1489.

\bibitem{2013_AMMR}
{\sc Amy, M., Maslov, D., Mosca, M., and Roetteler, M.}
\newblock A meet-in-the-middle algorithm for fast synthesis of depth-optimal
  quantum circuits.
\newblock {\em IEEE Transactions on Computer-Aided Design of Integrated
  Circuits and Systems 32}, 6 (2013), 818--830.

\bibitem{2015_BRS}
{\sc Bocharov, A., Roetteler, M., and Svore, K.~M.}
\newblock Efficient synthesis of universal repeat-until-success quantum
  circuits.
\newblock {\em Physical review letters 114}, 8 (2015), 080502.

\bibitem{2012_BAOKM}
{\sc Bombin, H., Andrist, R.~S., Ohzeki, M., Katzgraber, H.~G., and
  Mart{\'\i}n-Delgado, M.~A.}
\newblock Strong resilience of topological codes to depolarization.
\newblock {\em Physical Review X 2}, 2 (2012), 021004.

\bibitem{2005_BK}
{\sc Bravyi, S., and Kitaev, A.}
\newblock Universal quantum computation with ideal clifford gates and noisy
  ancillas.
\newblock {\em Physical Review A 71}, 2 (2005), 022316.

\bibitem{2012_BSK_}
{\sc Britton, J.~W., Sawyer, B.~C., Keith, A.~C., Wang, C.-C.~J., Freericks,
  J.~K., Uys, H., Biercuk, M.~J., and Bollinger, J.~J.}
\newblock Engineered two-dimensional ising interactions in a trapped-ion
  quantum simulator with hundreds of spins.
\newblock {\em Nature 484}, 7395 (2012), 489.

\bibitem{2011_BWC_}
{\sc Brown, K.~R., Wilson, A.~C., Colombe, Y., Ospelkaus, C., Meier, A.~M.,
  Knill, E., Leibfried, D., and Wineland, D.~J.}
\newblock Single-qubit-gate error below 10- 4 in a trapped ion.
\newblock {\em Physical Review A 84}, 3 (2011), 030303.

\bibitem{2012_CGC_}
{\sc Chow, J.~M., Gambetta, J.~M., C{\'o}rcoles, A., Merkel, S.~T., Smolin,
  J.~A., Rigetti, C., Poletto, S., Keefe, G.~A., Rothwell, M.~B., Rozen, J.,
  et~al.}
\newblock Universal quantum gate set approaching fault-tolerant thresholds with
  superconducting qubits.
\newblock {\em Physical review letters 109}, 6 (2012), 060501.

\bibitem{2019_dBBW}
{\sc de~Beaudrap, N., Bian, X., and Wang, Q.}
\newblock Techniques to reduce $\pi/4$-parity phase circuits, motivated by the
  zx calculus.
\newblock {\em arXiv preprint arXiv:1911.09039\/} (2019).

\bibitem{2016_dMM}
{\sc Di~Matteo, O., and Mosca, M.}
\newblock Parallelizing quantum circuit synthesis.
\newblock {\em Quantum Science and Technology 1}, 1 (2016), 015003.

\bibitem{2009_FSG}
{\sc Fowler, A.~G., Stephens, A.~M., and Groszkowski, P.}
\newblock High-threshold universal quantum computation on the surface code.
\newblock {\em Physical Review A 80}, 5 (2009), 052312.

\bibitem{2012_FWH}
{\sc Fowler, A.~G., Whiteside, A.~C., and Hollenberg, L.~C.}
\newblock Towards practical classical processing for the surface code.
\newblock {\em Physical review letters 108}, 18 (2012), 180501.

\bibitem{2013_GS}
{\sc Giles, B., and Selinger, P.}
\newblock Exact synthesis of multiqubit clifford+ t circuits.
\newblock {\em Physical Review A 87}, 3 (2013), 032332.

\bibitem{2013_GKMR}
{\sc Gosset, D., Kliuchnikov, V., Mosca, M., and Russo, V.}
\newblock An algorithm for the t-count.
\newblock {\em arXiv preprint arXiv:1308.4134\/} (2013).

\bibitem{1998_G}
{\sc Gottesman, D.}
\newblock The heisenberg representation of quantum computers.
\newblock {\em arXiv preprint quant-ph/9807006\/} (1998).

\bibitem{1999_GC}
{\sc Gottesman, D., and Chuang, I.~L.}
\newblock Quantum teleportation is a universal computational primitive.
\newblock {\em arXiv preprint quant-ph/9908010\/} (1999).

\bibitem{2013_J}
{\sc Jones, C.}
\newblock Low-overhead constructions for the fault-tolerant toffoli gate.
\newblock {\em Physical Review A 87}, 2 (2013), 022328.

\bibitem{2003_K}
{\sc Kitaev, A.~Y.}
\newblock Fault-tolerant quantum computation by anyons.
\newblock {\em Annals of Physics 303}, 1 (2003), 2--30.

\bibitem{2013_K}
{\sc Kliuchnikov, V.}
\newblock Synthesis of unitaries with clifford+ t circuits.
\newblock {\em arXiv preprint arXiv:1306.3200\/} (2013).

\bibitem{2012_KMM}
{\sc Kliuchnikov, V., Maslov, D., and Mosca, M.}
\newblock Fast and efficient exact synthesis of single qubit unitaries
  generated by clifford and t gates.
\newblock {\em arXiv preprint arXiv:1206.5236\/} (2012).

\bibitem{2013_KMM2}
{\sc Kliuchnikov, V., Maslov, D., and Mosca, M.}
\newblock Asymptotically optimal approximation of single qubit unitaries by
  clifford and t circuits using a constant number of ancillary qubits.
\newblock {\em Physical review letters 110}, 19 (2013), 190502.

\bibitem{2013_KMM}
{\sc Kliuchnikov, V., Maslov, D., and Mosca, M.}
\newblock Fast and efficient exact synthesis of single-qubit unitaries
  generated by clifford and t gates.
\newblock {\em Quantum Information \& Computation 13}, 7-8 (2013), 607--630.

\bibitem{2014_lG}
{\sc Le~Gall, F.}
\newblock Powers of tensors and fast matrix multiplication.
\newblock In {\em Proceedings of the 39th international symposium on symbolic
  and algebraic computation\/} (2014), ACM, pp.~296--303.

\bibitem{2016_M}
{\sc Maslov, D.}
\newblock Advantages of using relative-phase toffoli gates with an application
  to multiple control toffoli optimization.
\newblock {\em Physical Review A 93}, 2 (2016), 022311.

\bibitem{2013_PR}
{\sc Paetznick, A., and Reichardt, B.~W.}
\newblock Universal fault-tolerant quantum computation with only transversal
  gates and error correction.
\newblock {\em Physical review letters 111}, 9 (2013), 090505.

\bibitem{2014_PS}
{\sc Paetznick, A., and Svore, K.~M.}
\newblock Repeat-until-success: non-deterministic decomposition of single-qubit
  unitaries.
\newblock {\em Quantum Information \& Computation 14}, 15-16 (2014),
  1277--1301.

\bibitem{2012_RGP_}
{\sc Rigetti, C., Gambetta, J.~M., Poletto, S., Plourde, B., Chow, J.~M.,
  C{\'o}rcoles, A., Smolin, J.~A., Merkel, S.~T., Rozen, J., Keefe, G.~A.,
  et~al.}
\newblock Superconducting qubit in a waveguide cavity with a coherence time
  approaching 0.1 ms.
\newblock {\em Physical Review B 86}, 10 (2012), 100506.

\bibitem{2016_RS}
{\sc Ross, N.~J., and Selinger, P.}
\newblock Optimal ancilla-free clifford+ t approximation of z-rotations.
\newblock {\em Quantum Information \& Computation 16}, 11-12 (2016), 901--953.

\bibitem{2013_S}
{\sc Selinger, P.}
\newblock Quantum circuits of t-depth one.
\newblock {\em Physical Review A 87}, 4 (2013), 042302.

\bibitem{2015_S}
{\sc Selinger, P.}
\newblock Efficient clifford+ t approximation of single-qubit operators.
\newblock {\em Quantum Information \& Computation 15}, 1-2 (2015), 159--180.

\bibitem{1994_S}
{\sc Shor, P.~W.}
\newblock Algorithms for quantum computation: Discrete logarithms and
  factoring.
\newblock In {\em Proceedings 35th annual symposium on foundations of computer
  science\/} (1994), Ieee, pp.~124--134.

\bibitem{1999_S}
{\sc Shor, P.~W.}
\newblock Polynomial-time algorithms for prime factorization and discrete
  logarithms on a quantum computer.
\newblock {\em SIAM review 41}, 2 (1999), 303--332.

\end{thebibliography}
\end{document}